\documentclass[twocolumn]{aastex62}

\usepackage[utf8]{inputenc}
\usepackage{amsfonts}
\usepackage{amsmath}
\usepackage{bbold}
\usepackage{mathrsfs}
\usepackage{amssymb}
\usepackage{textgreek}
\usepackage[nolist,nohyperlinks,printonlyused]{acronym}

\usepackage[english]{babel}
\usepackage{fixme}
\usepackage{savesym}
\usepackage{verbatim}

\savesymbol{tablenum}
\usepackage{siunitx}
\restoresymbol{SIX}{tablenum}
\usepackage[english]{babel}

\FXRegisterAuthor{ar}{aar}{Add reference}
\bibpunct{(}{)}{;}{a}{}{,}

\newcommand{\vect}[1]{\boldsymbol{#1}}

\newcommand\given[1][]{\:#1\vert\:}
\newcommand{\lyaffirst}{\text{Lyman-$\alpha$ (Ly$\alpha$)}}
\newcommand{\lyaf}{\text{Ly$\alpha$ forest}}
\newcommand{\lya}{\text{Ly$\alpha$}}
\newcommand{\lyb}{\text{Ly$\beta$}}
\newcommand{\NHI}{$N_{\text{HI}}$}

\newcommand{\NHIt}{N_{\text{HI}} }
\newcommand{\bvel}{$b$}
\newcommand{\bnpdf}{$P ( \log N_{\text{HI}}, \log b \given \log T_0, \gamma )$}
\newcommand{\pbn}{$P(\log N_{\text{HI}}, \log b)$}
\newcommand{\bnpdfm}{P(\log N_{\text{HI}}, \log b \given \log T_0, \gamma)}
\newcommand{\logbnpdf}{$\ln P(\log N_{\text{HI}}, \log b \given \log T_0, \gamma) $}
\newcommand{\cmtwo}{\text{cm}$^{-2}$}
\newcommand{\vpfit}{\texttt{VPFIT}}
\newcommand{\bndist}{$b$-$N_{\text{HI}}$ distribution}

\DeclareSIUnit \parsec{pc}
\DeclareSIUnit \h {\ensuremath{\mathit{h}}}
\DeclareSIUnit\com{com. }
\DeclareSIUnit\yr{yr}

\newcommand{\kms}{\si{\kilo\meter\per\second}}

\newcommand{\Mpc}{\si{\mega\parsec}}
\newcommand{\kpc}{\si{\kilo\parsec}}

\newcommand{\lyatitle}{L\MakeLowercase{yα}}
\newcommand{\btitle}{\MakeLowercase{$b$}}

\renewcommand\plotone[1]{ 
 \centering 
 \leavevmode 
 \includegraphics[width={0.95\linewidth}]{#1} 
} 
\renewcommand\plottwo[2]{ 
 \centering 
 \leavevmode 
 \includegraphics[width={0.475\linewidth}]{#1} 
 \hfil 
 \includegraphics[width={0.475\linewidth}]{#2} 
}

\def\equationautorefname~#1\null{eqn.~(#1)\null}

\renewcommand{\textbf}{}
\renewcommand{\bfseries}{}

\begin{document}
\title{A Novel Statistical Method for Measuring the Temperature-Density Relation in the IGM\\ Using the $b$-$N_{\text{HI}}$ Distribution of absorbers in the \lyatitle{} Forest}

\author[0000-0003-4544-9437]{Hector Hiss}
\affiliation{Max-Planck-Institut für Astronomie, Königstuhl 17,
69117 Heidelberg, Germany}
\affiliation{International Max Planck Research School for Astronomy \&
Cosmic Physics at the University of Heidelberg}

\author[0000-0001-9182-6972]{Michael Walther}
\affiliation{Max-Planck-Institut für Astronomie, Königstuhl 17,
69117 Heidelberg, Germany}
\affiliation{International Max Planck Research School for Astronomy \&
Cosmic Physics at the University of Heidelberg}
\affiliation{Physics Department, Broida Hall, University of California Santa Barbara, Santa Barbara, CA
93106-9530, USA}

\author[0000-0002-8723-1180]{Jose O\~norbe}
\affiliation{Max-Planck-Institut für Astronomie, Königstuhl 17,
69117 Heidelberg, Germany} 
\affiliation{Royal Observatories, Blackford Hill, Edinburgh EH9 3HJ, UK}

\author[0000-0002-7054-4332]{Joseph F. Hennawi}
\affiliation{Max-Planck-Institut für Astronomie, Königstuhl 17,
69117 Heidelberg, Germany}
\affiliation{Physics Department, Broida Hall, University of California Santa Barbara, Santa Barbara, CA
93106-9530, USA}

\email{hiss@mpia.de}

%%%%%%%%%%%%%%%%%
%%%%%%%%%%%%%%%%%
\begin{abstract}%
%%%%%%%%%%%%%%%%%
%%%%%%%%%%%%%%%%% 
We present a new method for determining the thermal state of the intergalactic medium based on Voigt profile decomposition of the
\lya{} forest. The distribution of Doppler parameter and column density (\bndist{}) 
is sensitive to the temperature density relation $T=T_0(\rho\slash{\rho_0})^{\gamma-1}$, and previous work has inferred
$T_0$ and $\gamma$ by fitting its low-$b$ cutoff. This approach discards the majority of
available data, and is susceptible to systematics related to 
cutoff determination. We present a method that exploits
all information encoded in the \bndist{} by modeling its
entire shape. We apply kernel density estimation to discrete absorption lines to generate model probability density functions, then use principal component decomposition to create an emulator which can be evaluated anywhere in thermal parameter space. We introduce a Bayesian likelihood based on these models enabling parameter inference via Markov chain Monte Carlo. The method's robustness is tested by applying it to a large grid of thermal history simulations.  
By conducting 160 mock measurements we establish that our approach delivers unbiased estimates and valid uncertainties for a 
2D $(T_0, \gamma)$ measurement. Furthermore, we conduct a pilot study applying this
methodology to real observational data at $z=2$.
Using 200 absorbers, equivalent in pathlength to a single \lyaf{} spectrum, 
we measure $\log T_0 =4.092^{+0.050}_{-0.055}$ and $\gamma=1.49^{+0.073}_{-0.074}$ in
excellent agreement with cutoff fitting determinations using the same
data. Our method is far more sensitive than cutoff fitting, enabling measurements of $\log T_0$ and $\gamma$ with precision on $\log T_0$ ($\gamma$) nearly two (three) times higher for current dataset sizes.
\end{abstract}

%%%%%%%%%%%%%%%%%%%%%%%%%%%%%%%%%%%%%%%%%%%%%%%%%%%%%%%%%%%%%%%%%%%%%%%%%%%
%%%%%%%%%%%%%%%%%%%%%%%% BEGIN OF TEXT %%%%%%%%%%%%%%%%%%%%%%%%%%%%%%%%%%%%
%%%%%%%%%%%%%%%%%%%%%%%%%%%%%%%%%%%%%%%%%%%%%%%%%%%%%%%%%%%%%%%%%%%%%%%%%%%

%%%%%%%%%%%%%%%%%%%%%%%%%%%%%%%%%%%
%%%%%%%%%%%%% SECTION %%%%%%%%%%%%%
\section{Introduction}
\label{sec:introduction} 
The low density \ac{IGM} is the major reservoir of baryonic matter in the Universe. 
As the universe undergoes 
phase transitions, such as a global reionization process, the thermal state of the \ac{IGM} is changed. 
Thus precise measurements of the thermal history of the \ac{IGM} are key for our understanding of the details 
of reionization processes in the universe. 

The established picture concerning reionization is that the universe
undergoes two major phase transitions that change the thermal state of
the baryons. Firstly the reionization of hydrogen\footnote{Due to
comparable ionization thresholds, it is normally assumed that helium
is singly ionized (\ion{He}{1}$\rightarrow$\ion{He}{2}) along with
\ion{H}{1}.} (\ion{H}{1}$\rightarrow$\ion{H}{2}) which is believed
to be completed by redshift $z\sim 6$ \citep{mcgreer1}. This
reionization process is believed to be driven by the first galaxies
\citep{Faucher-Giguere2008,Robertson2015}, but it has recently been
debated whether early QSOs (quasi stellar objects, or quasars) could
have contributed substantially \citep{Madau_qso, Khaire16,Kulkarni2018}.

Once the population of luminous QSOs becomes abundant, there are enough
high energy photons available to power a second phase transition,
namely the second reionization of Helium
(\ion{He}{2}$\rightarrow$\ion{He}{3}) 
\citep[see e.g.][]{MadauMeiksin1994, MiraldaEscude2000, mcquinn09, Dixon2009, compostella1, compostella2, Syphers2014, Dixon2014}. 
Due to the requirement of a
large QSO population, this process becomes only possible at much later
times and is expected to be completed by $z\simeq2.7$
\citep[see e.g.][]{Worseck2011, Worseck2018}. Understanding the thermal imprint of
these processes is key for understanding the details of reionization
processes, i.e. their evolution and the sources powering them.

The main driving forces governing the thermal state of the \ac{IGM} (at $z\lesssim5$) 
are heating caused by photoionization by the \ac{UVB} 
and adiabatic cooling due to the expansion of the universe. 
It can be shown that long after the impulsive heating by reionization events \citep{mcquinn09, compostella1, mcquinn16}, the 
majority of the gas is naturally driven to a tight \ac{TDR} with the form $T = T_0 (\rho/\rho_0)^{\gamma-1}$ \citep{hui1},
where $T_0$ is the temperature at mean density $\rho_0$, and the power law index $\gamma$ quantifies
the temperature contrast between underdensities 
and overdensities. 

Since intergalactic gas is so diffuse, it is extremely challenging to study its properties in 
emission. 
Therefore, most of the knowledge we have about the \ac{IGM} comes from observing it in 
absorption. 
The primary observable at $z \lesssim 6$ that contains information about the thermal state of the \ac{IGM} is the \lyaffirst{} 
forest \citep{GunnPeterson1965, lynds}. 
This fluctuating absorption, consisting of a series of redshifted \lya{} absorption features in the 
lines of sight toward luminous objects (QSOs), arises from the fact that residual neutral hydrogen is present in the 
diffuse \ac{IGM}. The \lyaf{} can be used in different ways as a probe of the thermal state of the intergalactic gas. 
This includes various statistical measures such as of the power spectrum of the 
transmitted flux \cite[e.g.][]{Zaldarriaga1,McDonald2006, Walther2018a,Khaire2018,Walther2018,Boera2018}, 
the curvature statistic \citep{becker1,boera1}, the flux probability
distribution function \cite[e.g.][]{bolton2008, viel1, lee1}, 
as well as wavelet decompositions of the forest \cite[e.g.][]{theuns1,lidz1,Garzilli2012}.

In this study we use a method that treats the \lyaf{} as a superposition of multiple discrete 
absorption profiles \citep{schaye3, ricotti1, mcdonald1}, whereby each 
absorption profile is described by its position in redshift space, a Doppler parameter $b$ describing 
the absorption line width, and a column density \NHI{} that characterizes the density along the line of sight 
causing the absorption.
The thermal state is encoded in the absorption profiles, as thermal random motions in the absorbing gas contribute 
to the Doppler parameter. This is simply a result of blue and redshifting of the absorption 
wavelength due to Maxwell-Boltzmann velocity distributions in the gas. 
Additionally, \textbf{the broadening of absorption profiles is increased by the by differential Hubble flow across the 
spatial extent of the absorber, set by the pressure smoothing scale $\lambda_P$ \citep{HuiGnedin1998,schaye2,Peeples2010,Rorai2013,Kulkarni2015,rorai_science}. 
Peculiar velocity structure along the line of sight also contributes to the width of absorbers}. 

The conventional method for measuring thermal parameters using the
joint distribution of column densities and Doppler parameters
(\bndist{}) of absorbers in the \lyaf{} in a particular redshift
interval relies on the measurement of the thermal state dependent
lower cutoff in this distribution \citep[see][]{schaye3, ricotti1,
mcdonald1, schaye1, rudie1, bolton1, Garzilli2015, Rorai17vpfit, Hiss2018, Telikova2018, 
Garzilli2018}, set primarily by the minimal broadening associated
with the temperature of the absorbers.

Although it constitutes a powerful tool for measuring the thermal state of the gas, 
the cutoff fitting technique has a series of inherent disadvantages. The main one being that 
the position of the cutoff is fitted using an iterative technique which excludes absorbers from the 
distribution. This means that a small number of absorbers is effectively used for measuring the position of the cutoff, resulting in 
diminished sensitivity of the method on the total number of absorbers 
in the dataset once the distribution is well populated \citep{schaye3}. 
In addition, narrow metal line absorbers, which are difficult to completely
identify and mask, can result in significant contamination around the cutoff,
compromising the precision with which the cutoff can be determined, and
adding systematics which are difficult to control. 
Another complication of this method, as shown in \citet{Hiss2018} 
in the context of the comparison with the results by \citet{Rorai17vpfit}, 
is that choice of cutoff fitting method (i.e. least-squares or mean-deviation minimization)
can lead to significantly different $T_0$ and $\gamma$ measurements. 
All of these problems call for a new method for interpreting the information about the thermal 
state of the \ac{IGM} encoded in the \bndist{}.

In this work we introduce, test, and apply a new method for constraining $T_0$ and $\gamma$ using 
the \bndist{}. The main difference with the traditional cutoff fitting approach is that we model the entire
distribution, and thus bypass the complications associated with quantifying the position of a lower cutoff. 
\textbf{While other studies employed a parametric description of the full \bndist{} in order to 
carry out measurements of the parameters of the \ac{TDR} \citep[see e.g.][]{ricotti1, Telikova2018}, 
we instead} construct smooth probability density functions (PDF) of simulated \bndist s 
using a non-parametric approach. These PDFs can then be used as models for conducting inference. 
\textbf{The reader should keep in mind that all results presented in our proof of concept 
concern $T_0$ and $\gamma$ alone and do not marginalize over other parameters. 
All results presented should be interpreted as a demonstration of the capabilities of this new approach 
rather than a perfect measurement.} 

This paper is structured as follows. We introduce our simulations and mock data generation in \S~\ref{sec:simulations}. 
Our new method for \textbf{constructing} a model of the
\bndist{} and inferring thermal parameters is described in \S~\ref{sec:method}. 
In \S~\ref{sec:mock_measurement}, we carry out measurements using different mock data realizations at $z=2$ to explore the robustness 
of this technique. 
We carry out a pilot study of this new method in \S~\ref{sec:actual_measurement}, where real observational
data at $z=2$ is compared to a grid of hydrodynamical simulations.
We discuss and summarize our results in \S~\ref{sec:summary}.

%%%%%%%%%%%%%%%%%%%%%%%%%%%%%%%%%%%
%%%%%%%%%%%%% SECTION %%%%%%%%%%%%%
\section{Simulations}
\label{sec:simulations}
In this section we describe how we generate simulated \lyaf{} spectra with different
combinations of the underlying thermal parameters that govern the IGM. Specifically, we wish to generate 
a grid of $T_0, \gamma$ at a fixed $\lambda_P$ to understand how the corresponding shape of the \bndist{} changes 
as a function of the thermal parameters $T_0, \gamma$, i.e. \bnpdf{}. Certainly the choice of $\lambda_P$ has an effect on the shape of 
the \bndist{}, as shown in \citet{Garzilli2015}, meaning that one should consider 
$P(\log \NHIt, \log b \given T_0, \gamma, \lambda_P)$. 
For the sake of simplifying the analysis for an initial proof of concept, 
we will test our method at a fixed $\lambda_P$. 
\textbf{Note that all cosmological length scales in this work are given in comoving units. }

For generating our $T_0, \gamma$ grid, we create 
mock spectra using a snapshot of a \ac{DM} only simulation at $z=2$.
Although it is well known that spectra \textbf{based on approximations to a full hydrodynamic 
simulation are limited in their ability
to accurately represent the IGM \citep{Gnedin1998,Meiksin2001,Viel2006,Sorini2016}, }
we opt to use \ac{DM} only simulations first, as they allow us to run many different thermal models in
a computationally feasible time, allowing us to generate dense thermal grids. 
This approach should suffice for initial tests, \textbf{as both mock data and models are generated from the 
same sort of simulation and} we are mainly interested 
in generating a method that is sensitive to thermal state dependent changes in the shape of the \bndist{}. 
We expand our analysis with the use of hydrodynamical simulations in \S~\ref{sec:actual_measurement}, which is a necessary 
step when dealing with actual observational data. 

Our simulation provides the dark-matter density and velocity fields 
calculated using an updated version of the TreePM code from \citet{White02} that evolves 
$N_p = 2048^3$ collisionless, equal mass particles ($M_p =2.5 \times 10^5 \text{M}_{\odot}$) in a periodic cube of side length
$L_{\text{box}} = 30 \, \Mpc /h$ 
with a Plummer equivalent smoothing of $1.2\,\kpc/h$ \citep[similar to][]{rorai1}. 
The cosmology used in the simulations is consistent within 1$\sigma$ with the 2013 Planck release \citep{planck} with 
$\Omega_{\Lambda}=0.691$, $\Omega_m=0.309$, $\sigma_8 = 0.829$, 
$\Omega_b h^2 = 0.022$, $n_s = 0.961$ and $h=0.678$. 

In order to model lines-of-sight through the \ac{IGM}, we extract skewers from our simulation that run parallel to one of the box axes 
and apply the recipe described below. 
A pseudo-baryonic field is generated by smoothing the dark-matter density and velocity fields. 
This smoothing mimics the effect of Jeans pressure smoothing of the gas, i.e. accounts for the fact that 
small-scale structure is suppressed in the baryonic matter distribution due to finite gas pressure 
\citep{Gnedin96, HuiGnedin1998, Kulkarni2015}.
We choose to smooth the dark-matter field with a 
constant (instantaneous) filtering scale $\lambda_P$. 
This is done by convolving the density and velocity fields in real-space with a cubic spline kernel of the form:
\begin{equation}
K(r,R_P) = \frac{8}{\pi R_P^3}
\begin{cases}
1-6 \left(\frac{r}{R_P}\right)^2 + 6 \left(\frac{r}{R_P}\right)^3 & \frac{r}{R_P} \leq \frac{1}{2} \\
2\left( 1-\frac{r}{R_P}\right)^3 &  \frac{1}{2} < \frac{r}{R_P} \leq 1\\
0 & \frac{r}{R_P} >1
\end{cases}
\end{equation}
with a smoothing parameter $R_P$. 
This function closely resembles a Gaussian with $\sigma \sim R_P / 3.25$ in the central regions, which defines our pressure 
smoothing scale $\lambda_P = R_P / 3.25$. 
Given the characteristics of our simulations, the mean inter-particle separation $\Delta \ell = L_{\text{box}}/N_p^{1/3}$ allows 
us to resolve values of $\lambda_P \gtrsim 20\,\kpc$ \citep{rorai1}. 
For all \ac{DM} only related models used in this work, we will adopt a fixed value of $\lambda_P = 73.3\,\kpc$, 
which is consistent with the measurement by \citet{rorai1} at $z=2$.

Under the assumption that the \ac{IGM} is highly ionized and in photoionization equilibrium, 
we can construct a \lya{} optical depth field in real space based on the smoothed dark matter density field using 
the fluctuating Gunn-Peterson approximation \citep[FGPA][]{Weinberg1997,fgpa} 
\begin{equation}
\tau(x) \propto n_{\text{HI}}(x) \propto T_0^{-0.7}\rho(x)^{2-0.7(\gamma-1)}, 
\end{equation}
where $x$ is the particle position in real space. 
In order to account for the effects of thermal broadening and peculiar velocities of the gas on the optical depth, 
we compute the redshift-space optical depth by convolving the real space optical depth with a Gaussian-profile. 
This is an approximation to the actual Voigt-profile and is characterized by a thermal width $b = \sqrt{2 k_B T /m_{\rm HI}}$, 
(where $m_{\rm HI}$ is the hydrogen atom mass, $k_b$ the Boltzmann constant and T the temperature) 
and a shift from its real-space position by the longitudinal component of the peculiar velocity. 
This way we can impose a deterministic power law \ac{TDR} onto the simulation, i.e. choose $T_0$ and $\gamma$.
This allows us to generate mock spectra with different sets of underlying thermal parameters $T_0$ and $\gamma$.

%%%%%%%%%%%%% FIGURE %%%%%%%%%%%%%
\begin{figure}
\plotone{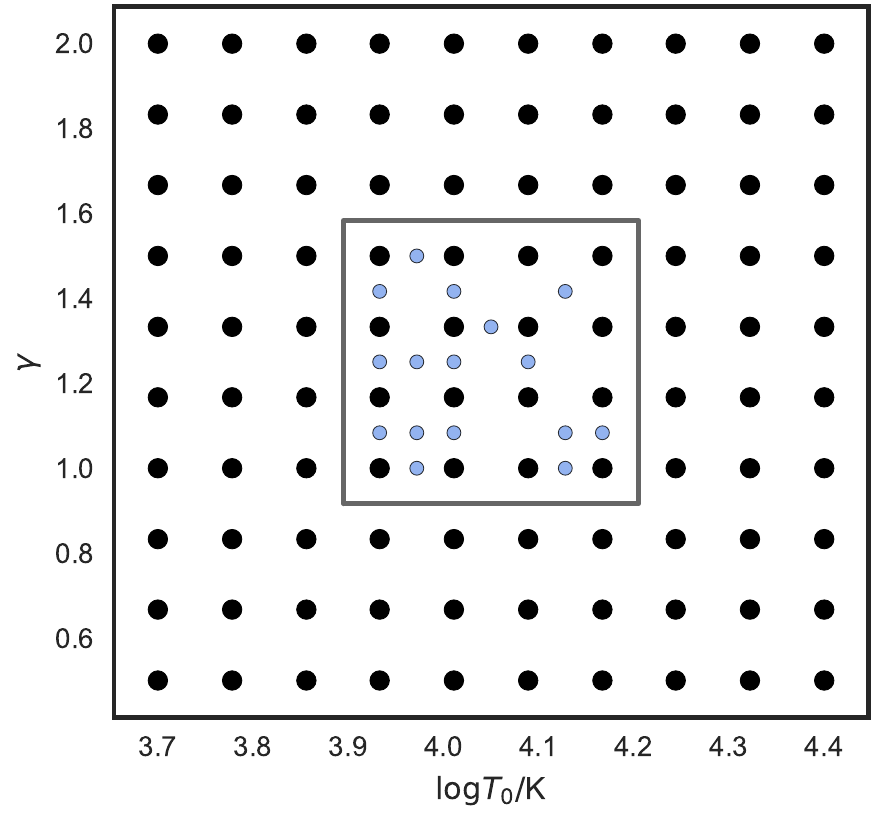}
\caption{Grid of thermal parameters applied to a \ac{DM} only simulation at $z=2$ used to construct a model of the 
\bndist{}. The black points show the combinations of $\log T_0$ and $\gamma$ imposed onto our simulation 
(``standard grid''). 
The square marks the area that will be used for inference tests. The blue points indicate where further models were generated 
for testing the robustness of the method presented in this work (``test grid'').} 
\label{fig:thermal_grid}
\end{figure} 
The corresponding flux skewer $F$, i.e. a transmission spectrum along the line-of-sight, is calculated 
from the optical depth using $F = \exp(-A_r \, \tau)$. Here we introduce a scaling factor $A_r$ that allows us to match our 
lines-of-sight to observed mean flux values $\bar{F}$. 
\textbf{The mean flux normalization is computed for the full snapshot, i.e. the factor $A_r$ is iteratively changed until 
the mean flux of the snapshot converges to a desired (measured) mean flux.
We apply that value of $A_r$ to all the spectra when generating skewers, so
there is one mean flux normalization of the whole box and sightline to sightline variations are still present in our models.} 
This re-scaling of the optical depth accounts for our lack of knowledge of the precise value
of the metagalactic ionizing background photoionization rate and it is done simply to generate more realistic skewers. 
To this end we choose $A_r$ so that we agree with the effective opacity 
$\tau_{\text{eff}} = -\ln (\bar{F})$ at $z=2$ from \citet{Faucher-Giguere2008b}, namely $\tau_{\text{eff}}=0.127$.

%%%%%%%%%%%%% FIGURE %%%%%%%%%%%%%
\begin{figure*}
 \IfFileExists{KDE_bw_0.08_0.032.pdf}{\plotone{./{KDE_bw_0.08_0.032}.pdf}}{\plotone{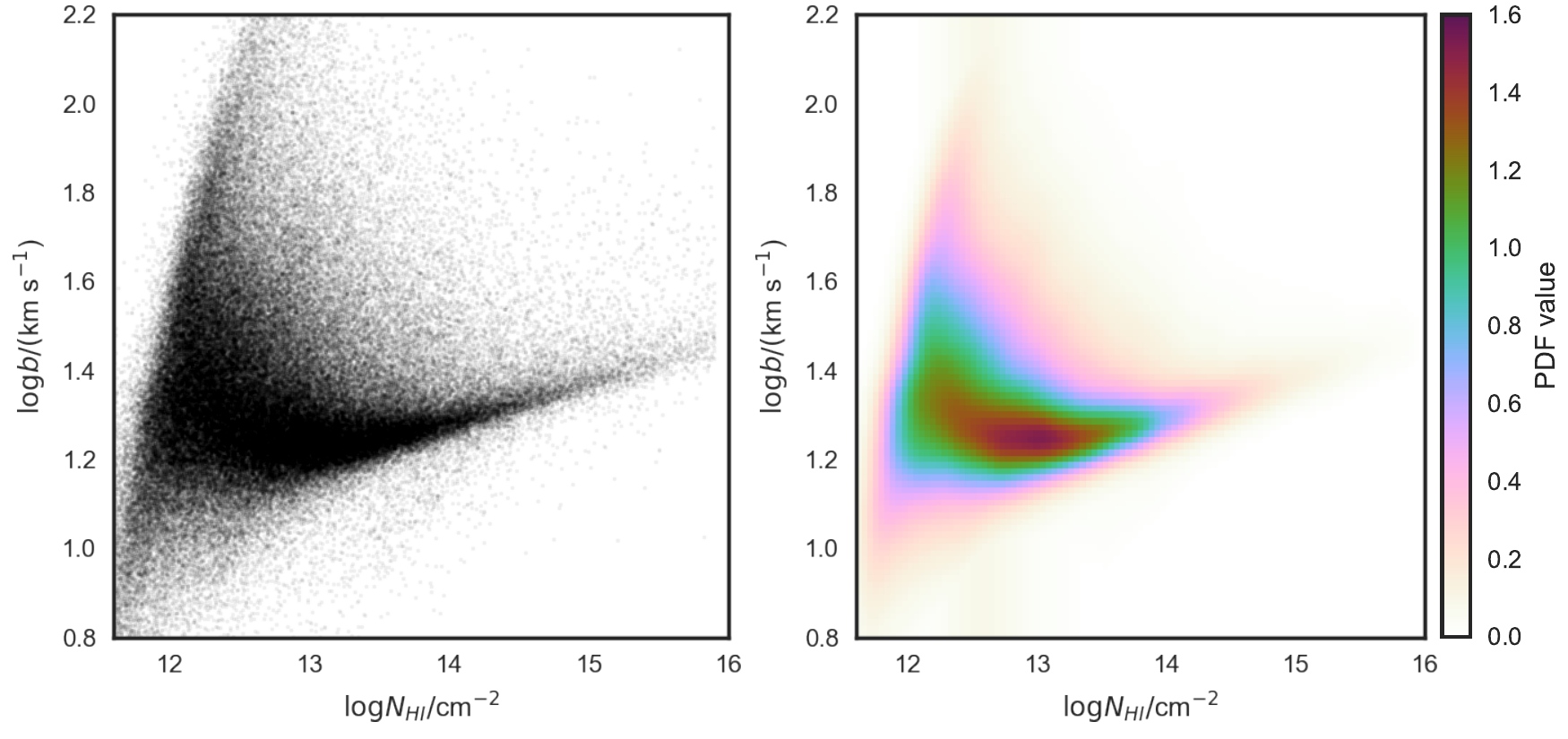}}
  \caption{\textbf{Left}: A \bndist{} illustrated as a cloud of points generated by 
  concatenating the \vpfit{} output for 6000 skewers from a DM only simulation snapshot at $z=2$ 
  with thermal parameters $(\log T_0, \gamma) = (4.011, 1.333)$. 
  This distribution consists of $\sim 1.5 \times 10^5$ absorbers. 
  \textbf{Right}: The KDE based \ac{PDF} of the same distribution 
  (as described in \S~\ref{subsec:kde}). } 
 \label{fig:kde_points}
\end{figure*}

%%% SUBSECTION %%%
\subsection{Thermal Parameter Grid}
\label{subsec:parameter_grid}
Using our simulation snapshot at $z=2$ we generated 6000 skewers 
for each of 100 combinations of thermal parameters $\log T_0$ and $\gamma$ at a fixed $\lambda_P = 73.3\,\kpc$. 
Figure~\ref{fig:thermal_grid} shows the distribution of thermal parameters chosen (black points). 
We chose to model the thermal parameters on a 10$\times$10 regular grid covering the range 
$ 3.7 \leq \log (T_0/\text{K}) \leq 4.4$ and $0.5 \leq \gamma \leq 2.0$, which is dense enough to sample 
typical uncertainties in $T_0$ and $\gamma$.
The number of skewers at each grid point was chosen, 
so that we have enough absorbers to ensure that our estimation of the shape of the \bndist{} 
is converged. This is important, as we will use the absorbers in the 
\bndist{} to estimate \bnpdf{} which we will introduce in \S~\ref{subsec:kde}. 
In this work we will refer to this grid as the ``standard grid''. 

In addition, we generated 16 models between the grid points in the central region of our grid 
(region marked with the square and blue points in Figure~\ref{fig:thermal_grid}). These were randomly chosen from a regular 
grid twice as fine as 
the standard grid, excluding the points that coincide with it. 
These additional models will be used in \S~\ref{subsec:emulator} to test the robustness of our procedure
for generating model \bndist s, as well as our statistical inference (see \S~\ref{subsec:inference_test}).
We will refer to these extra models as the ``test grid''.

%%% SUBSECTION %%%
\subsection{Forward Modeling Noise and Resolution}
\label{subsec:forward_modeling} 
The technique presented in the following section is based on the sensitivity of 
the shape of the \bndist{} on the thermal state of the \ac{IGM}. 
Therefore, it is important that instrumental effects which can also affect the shape of the \bndist{}, such as noise and spectroscopic resolution, 
are properly included into the models we wish to compare to data. 

To mimic instrumental resolution we convolve 
the skewers with a Gaussian with FWHM = 6\,\kms, which is the typical resolution delivered by echelle spectrometers
\citep[see e.g. the \ac{HIRES} \citep{hires,kodiaq1,kodiaqdr1,kodiaqDR2} and \ac{UVES} \citep{dekker, aldo1} dataset in][]{Hiss2018}. 
Further, we add Gaussian random noise to the skewers assuming a fixed \ac{SNR} 
of 63 per resolution element for the purpose of choosing a value 
comparable to the \ac{SNR} of the dataset in \citet{Hiss2018} at $z=2$. 

We apply the exact same Voigt-profile fitting scheme described in \citet{Hiss2018} to the 6000 forward modeled simulated skewers 
generated for 100 different combinations of $T_0$, $\gamma$. 
To summarize, Voigt-profiles were fitted to our simulated data using \vpfit{} version 10.2\footnote{VPFIT: \url{http://www.ast.cam.ac.uk/~rfc/vpfit.html}} 
\citep{vpfit}.
We wrote a fully automated set of wrapper 
routines that prepare the spectra for the fitting procedure and 
controls \vpfit{} with the help of the \vpfit{} front-end/back-end programs \texttt{RDGEN} and 
\texttt{AUTOVPIN}. 

\vpfit{} decomposes segments of spectra into a set of Voigt-profiles characterized by 3 parameters each: 
line redshift $z_{\text{abs}}$, Doppler parameter $b$, and column density \NHI{} 
for the Hydrogen \lya{} transition. 
We set up \vpfit{} to explore the range of parameters 
$1 \leq b/\kms \leq 300$ and $11.5 \leq \log(\NHIt{}/\text{\cmtwo{}}) \leq 16$ when fitting absorption profiles. 
\textbf{We chose to fit in this \NHI{} range in order 
encompass typical optically thin \lya{} absorbers ranging from low column densities 
(where most of lines are comparable to noise) to very rare high column densities. 
Concerning the Doppler parameter, the chosen fitting region ranges from narrow absorbers, that are unphysical and have broadening comparable to the \ac{UVES}/\ac{HIRES} resolution element, 
to broad absorbers that are substantially broader than the typical absorber around the cutoff for all 
$\log T_0$ and $\gamma$ combinations in our grid. 
This choice of fitting range is appropriate, as the probability of encountering absorbers close to the edge of our fitting range 
drops to nearly zero at this redshift.}

\textbf{\vpfit{} finds the best fit by varying the profile} parameters and searching for a solution that minimizes the $\chi^2$. 
If the $\chi^2$ is not satisfying, then further absorption components are added until the fit converges or no longer improves. 
We take into account that \vpfit{} often has difficulty fitting the
boundaries of spectra by artificially increasing the length of the sightlines. For this purpose we append the first (last) 
quarter of the spectra to the end (beginning) of it, therefore making the spectra longer by 50\%. This manipulation does not cause 
discontinuities in the flux, as the simulation box is periodic. 
We later ignore absorbers within the artificially enlarged areas.

Additionally, in order to avoid using badly constrained absorber parameters, we exclude points that have relative 
uncertainties worse than 50\% in $b$ or \NHI{}. 
\textbf{These lines are rejected in order to remove absorbers that are badly constrained. 
As discussed in \citet{rudie1} and \citet{Hiss2018}, most of these lines arise in blended and noisy regions. 
Additionally, as the $\log$ errors are proportional to the relative errors, we expect a 50\% relative error to be $\text{d} \log x = \ln(10) \cdot \text{d} x / x = \ln(10) \cdot 0.5 \simeq 1.15$ (x being either \NHI{} or $b$). 
These uncertainties are substantially larger than our kernel density estimation bandwidth used in this study (see \S~\ref{subsec:kde}) which 
additionally motivated us to to exclude these absorbers. 
Finally, filtering these lines consistently in data and models should not bias our results, as these are mostly VPFIT artifacts and 
will consistently arise whenever there is noise and blending.}

For every combination of $\log T_0$ and $\gamma$,
a \bndist{} can be generated from all absorbers found 
for all skewers. One example with $(\log T_0, \gamma) = (4.011, 1.333)$ is shown in the left panel of Figure~\ref{fig:kde_points}. 

%%%%%%%%%%%%%%%%%%%%%%%%%%%%%%%%%%%
%%%%%%%%%%%%% SECTION %%%%%%%%%%%%%
\section{Method for Emulating the Full \btitle{}-$N_{\text{HI}}$ Distribution}
\label{sec:method}
In this section, we introduce the method used to generate \ac{PDF}s of \bndist s 
at any location in thermal parameter space based on our grid of simulated thermal models. For
each thermal model, we perform \ac{KDE} to determine \pbn{} 
from the discrete absorbers identified by \vpfit{}. 
To interpolate the \bndist{} between points in our parameter grid we modified the emulation technique 
of \citet{heitmann2006Cosmiccalibration} and \citet{habib2007CosmicCalibrationConstraints}, initially developed for power spectrum analysis, to our 
purpose. Note that this approach has also been used in the context of measurements of the evolution of the thermal state of the IGM 
in \citet{rorai1,rorai_science} and \cite{Walther2018}. 

We apply \ac{PCA} 
to decompose this set of \textbf{probability distribution maps} onto a set of basis vectors, yielding a set of coefficients
$\Theta_j(T_0,\gamma)$ for each thermal model corresponding to principal component vectors $\vect{e}_j$. 
We then use Gaussian process (GP) interpolation
to evaluate these coefficients at arbitrary locations in parameter space, which combined with 
the basis vectors, results in a model for \bnpdf{}.

Finally, we present a Bayesian method for determining the posterior
distribution of thermal parameters from an observed set of $\log \NHIt$ and $\log b$.
We refer to this procedure of model construction and inference, based on \ac{PCA} decomposition of \ac{KDE} estimates of a \ac{PDF}, as the PKP method. 
The details of each step are discussed in what follows. 

%%% SUBSECTION %%%
\subsection{Kernel density estimation of the \bndist{} \ac{PDF}}
\label{subsec:kde} 
In the first step of the PKP approach we use \ac{KDE} 
to construct the probability density distribution 
from which points in the \bndist s of our models 
were drawn.
This is achieved by treating each data point $\{\log N_{\rm{HI},i}, \log b_i\}$ as a smooth kernel centered at the measurements position 
$\log N_{\rm{HI},i}$ and $\log b_i$. 
We use a Gaussian kernel of the form 
\begin{align}
 K_i &(\sigma_{\log \NHIt}, \sigma_{\log b}) = \frac{1}{2 \pi  \sigma_{\log N_{\rm{HI}}} \sigma_{\log b}} \times  \label{eq:kernel}\\
 &\exp \left({-\frac{1}{2} \left[\frac{(\log N_{\rm{HI}} - \log N_{\rm{HI},i})^2}{\sigma_{\log N_{\rm{HI}}}^2}+\frac{(\log b - \log b_i)^2}{\sigma_{\log b}^2} \right]} \right), \nonumber
\end{align}
characterized by a bandwidth $(\sigma_{\log \NHIt}, \sigma_{\log b})$ that regulates how much one wishes to smooth a measurement 
in each dimension. Note that the Kernel used in eqn.~\ref{eq:kernel} assumes no correlation between 
$\log N_{\rm{HI},i}$ and $\log b_i$ for a given pair. This assumption should not significantly affect the estimated PDFs, because the single 
Kernels overlap substantially. 

With every measurement described as a smooth distribution, we can generate an estimate for 
the probability density function from which a set of measurements $\{\log N_{\rm{HI},j}, \log b_j\}$ with $j=1,...,N$, was drawn
\begin{equation}
 P(\log N_{\rm{HI}}, \log b) = \frac{1}{N} \sum_{j=1}^{N} K_j(\sigma_{\log \NHIt}, \sigma_{\log b}).
 \label{eq:kde}
\end{equation}
In other words, we compute \pbn{} by replacing each measurement with a Gaussian kernel with a constant bandwidth, summing them up 
and normalizing the distribution. 

%%%%%%%%%%%%% FIGURE %%%%%%%%%%%%%
\begin{figure*}
 \plotone{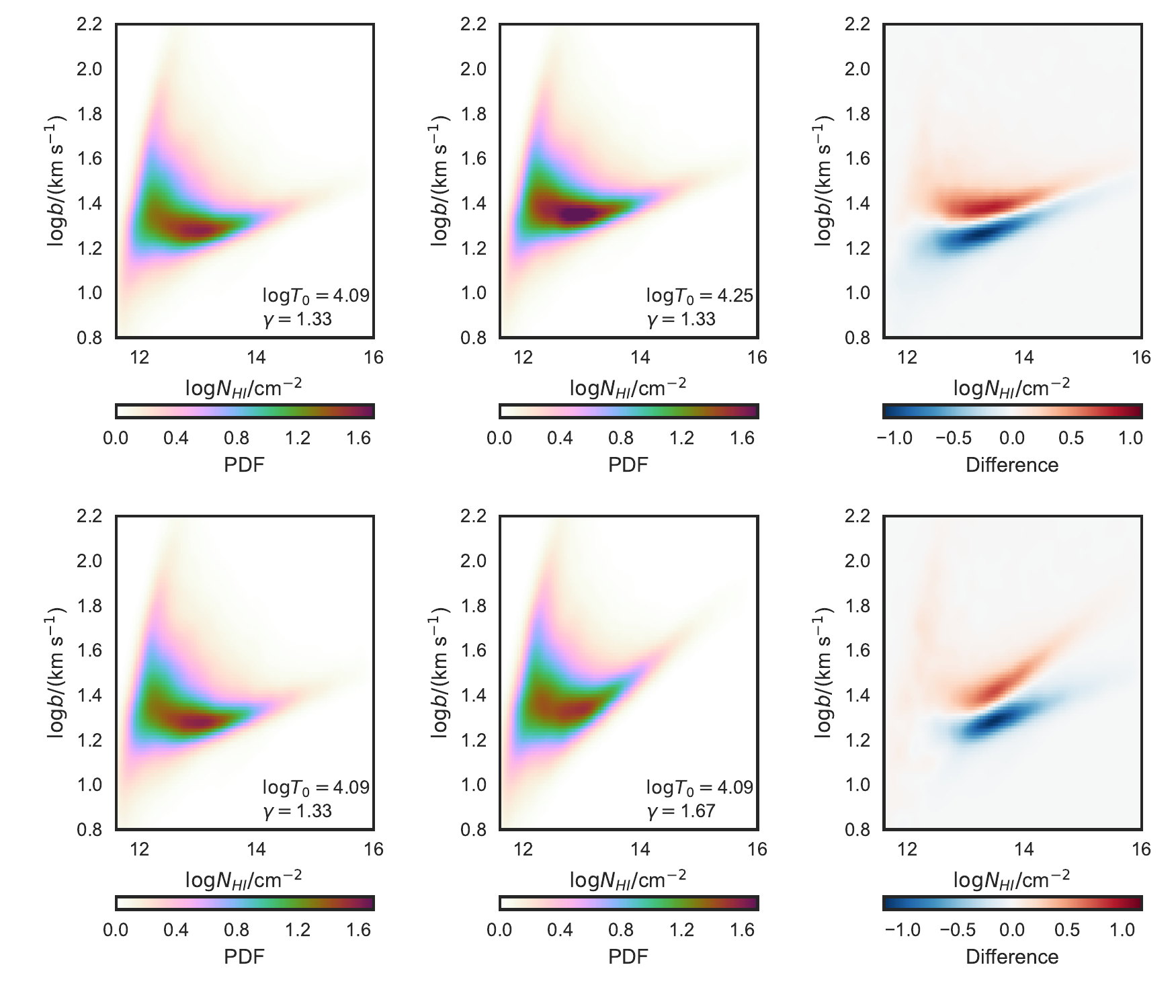}
  \caption{Sensitivity of \pbn{} on the thermal parameters imposed onto our simulation. 
  \textbf{Left:} \pbn{} based on the KDE of \bndist s at a given point in our thermal grid. 
  \textbf{Middle:} \pbn{} based on the KDE of \bndist s at a given point in our thermal grid choosing higher $\log T_0$ 
  (upper panel) and higher $\gamma$ (lower panel). 
  \textbf{Right:} The difference between the distributions illustrates that increasing $\log T_0$ shifts \pbn{} toward higher $b$ (top), while 
  increasing $\gamma$ mainly tilts the distribution at $\log (\NHIt$/\cmtwo{})$>13$ (bottom).}
 \label{fig:thermal_sensitivity}
\end{figure*}

In this study, we compute KDEs using the package {\tt KDEMultivariate} from the {\tt statsmodels} python 
module \citep{statsmodels}. 
An example of this method applied to one of our \bndist s is shown in the right panel of Figure~\ref{fig:kde_points} for one particular 
combination of thermal parameters $(\log T_0, \gamma) = (4.011, 1.333)$, which can be compared to the points in the 
\bndist{} determined by \vpfit{} in the left panel.

We generate \ac{KDE} based \pbn{} for every thermal parameter
combination in our standard thermal grid by applying \ac{KDE} to the
points in the \bndist{} determined by \vpfit, using a bandwidth of
$(\sigma_{\log N_{\rm HI}}, \sigma_{\log b} )= (0.08, 0.032)$ for each dimension. 
We tuned our bandwidth using mock datasets in order avoid oversmoothing of 
\bnpdf{}, which can wash out structure in the distribution. Additionally, oversmoothing shifts
the peak of \bnpdf{} towards high $b$ due to the asymmetry of the
distribution, resulting in a distribution that has its maximum clearly
shifted from the highest concentration of absorbers in the cloud of
points used to generate it. At the same time we were careful not to
undersmooth the distribution, which leads to a noisy \ac{PDF}.

For comparison, a Silverman estimation of the optimal bandwidth \citep{Silverman86} for our dataset, which assumes that 
the underlying distribution is Gaussian, typically yields 
a bandwidth of $(0.1, 0.04)$. This choice resulted in a very slight bias in our measurements
for mock data in the context of the inference test described in \S~\ref{subsec:inference_test},
indicating that this choice of bandwidth oversmoothes our distributions. 

To illustrate the sensitivity of our \ac{PDF} to thermal parameters we show \bnpdf{} for different 
$\log T_0$ and $\gamma$ combinations in Figure~\ref{fig:thermal_sensitivity}. 
\textbf{We observe that, as expected, most of the sensitivity 
of the \bndist{} with respect to the parameters of the \ac{TDR} lies in its lower $b$ envelope. Therefore, in the limit of a measurement of $T_0$ and $\gamma$,  
our approach can be interpreted as an alternative way of retrieving the cutoff 
(although without many of the problems associated with iterative cutoff fitting as described in \S~\ref{sec:introduction}). 
Nevertheless, our method can be expanded to any 
changes in the general form of the \bndist{}, provided that these are properly modeled in the simulations. 
The example of $T_0$ and $\gamma$ is an interesting starting point to apply our method to, but should not be seen as its sole application.
We know for instance that $\lambda_P$ \citep{Garzilli2015, Garzilli2018}, the fraction of the gas in the 
warm-hot phase \citep{Danforth2016} and galactic feedback \citep{Viel2017} affect the shape of the \bndist{} 
above the location of the cutoff. In principle, our method should be sensitive to these parameters as well.} 

For better intuition about the 
thermal sensitivity of the \bndist{} we also added Figures, constructed from the output of 
hydrodynamical simulations described in 
\S~\ref{subsec:hydro_models}, to appendix~\ref{app:animations}. 
These can be be viewed as animations in the HTML version of this manuscript (available in the refereed version only). 

%%% SUBSECTION %%%
\subsection{Decomposition of the PDF into Principal Components}
\label{subsec:pca_decomposition}

Given the non-parametric nature of \ac{KDE}, there is no direct way to generate \bnpdf{} 
for combinations of $\log T_0$ and $\gamma$ between points in our thermal grid positions. 
For this to be possible, we have to parametrize the \bnpdf{} maps. 
To this end, we evaluate the KDE of each \bndist{} on a $100\times100$ mesh
in the $b$-$N_{\rm HI}$ plane
and then decompose these pixelized \ac{PDF}s onto a set of linear independent principal components, 
thus parametrizing the \ac{KDE} based \bnpdf{} with \ac{PCA} coefficients and a set of basis vectors. 

Specifically, we discretized the PDFs in the region $11.5\leq \log (\NHIt / \text{\cmtwo}) \leq 16.$ 
and $0.8 \leq \log (b/\kms) \leq 2.2$), adopting a pixel size $(0.04, 0.014)$ in $(\log \NHIt, \log b)$, 
which is a factor of 2 smaller than the bandwidth chosen for the
\ac{KDE}. Then we compute the (natural) logarithm of the probabilities at every pixel.
Given our small pixels, we expect no significant change in the shape of the
\bndist{} due to pixelization. All examples of smooth
\bndist s shown in this work are pixelized on this grid (see e.g. Figures~\ref{fig:kde_points}, 
\ref{fig:thermal_sensitivity}, and \ref {fig:int_KDE}). 

%%%%%%%%%%%%% FIGURE %%%%%%%%%%%%%
\begin{figure*}
 \plotone{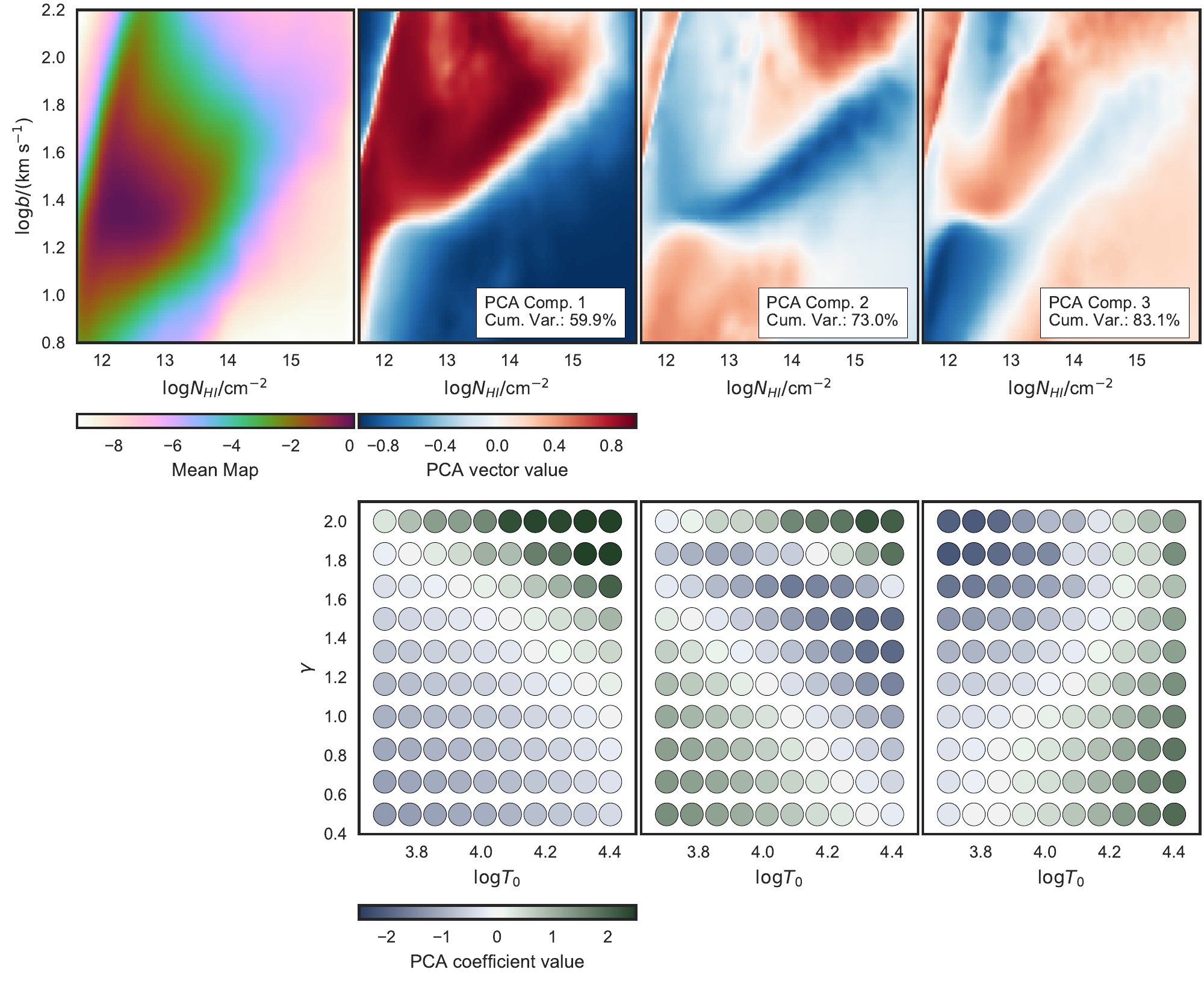}
  \caption{\textbf{Upper row:} The mean map $\vect{\mu}(\log N_{\text{HI}}$, $\log b$) and the first 3 principal component vectors $\vect{e}_j$ from our principal 
  component analysis 
  of our model maps. Note that the decomposition was carried out in the natural logarithm of the probability. The vectors were reshaped to the map 
  form of $100\times100$ pixels and are sorted by contribution to the 
  cumulative variance (see text for details). \textbf{Lower row:} The corresponding principal component 
  coefficients $\Theta_j(\log T_0, \gamma)$ for each map.}
  \label{fig:pca_components}
\end{figure*}
The \ac{PCA} is performed by decomposing our discrete maps into a basis of principal component vectors $\vect{e}_j$, which 
makes it possible to recover any model in our grid by linearly combining the principal component vectors, 
using the coefficients $\Theta_j(\log T_0,\gamma)$ and adding them to the mean map $\vect{\mu}(N_{\text{HI}}, b)$:
\begin{align}
 \ln \bnpdfm  = \vect{\mu}(\log N_{\text{HI}}, \log b) \\ \nonumber
              +\sum_{j=1}^N  \Theta_j(\log T_0, \gamma) \vect{e}_j(\log N_{\text{HI}}, \log b), 
\end{align}
where $N$ is the number of models available, in this case $N=100$, and the components are ranked by their contribution
to the cumulative variance of the dataset. 
In short, the PCA decomposes a matrix of all vectorized $\ln$\pbn{} maps into a basis of 100 principal component vectors with 
100 coefficients each.

In Figure~\ref{fig:pca_components} we show the $\vect{\mu}(\log N_{\text{HI}}, \log b)$ map and the first 3 principal component 
\textbf{vectors (reshaped to an image of 100$\times$100 pixels) and coefficients} from our analysis. 
Note that \ac{PCA} is a standard method for dimensionality reduction, as it allows one to choose the 
principal components that encompass most of the variance within the data by ignoring 
components that do not contribute substantially to the cumulative variance. The cumulative contribution 
to the total variance is computed by first dividing the eigenvalues from the singular value decomposition 
method used in the \ac{PCA} by their sum, ordering them in descending order, and computing their 
cumulative sum. 
For illustration, the first 3 components shown in Figure~\ref{fig:pca_components} already account for 83.1\% of the cumulative
variance in the models.
At present, we are not interested in dimensionality reduction and keeping all 100 \ac{PCA} components is 
not computationally prohibitive for the current case. 
By \ac{PCA} decomposing the \ac{KDE}s in our grid, we are simply describing each of the discretized 
\bnpdf{} with a set of coefficients $\Theta_j(T_0,\gamma)$ and basis vectors, enabling a parametric description of
\bnpdf{}. 

There are two reasons why we carried out the \ac{PCA} on $\ln$\pbn{}. 
First, because we will interpolate \ac{PCA} components of \logbnpdf{} maps (\S~\ref{subsec:emulator})
in thermal parameter space, and these PDFs have sharp features (such as the low $b$ cutoff). \textbf{Computing} the natural 
logarithm is desirable to reduce interpolation errors. 
Second, we do this for a practical reason, as we will ultimately tie this 
analysis to a \ac{MCMC} algorithm that works with the $\log$-likelihood. 

The disadvantage of working with the natural logarithm of 
\bnpdf{} is that the probability fluctuations around zero are amplified, which can destabilize the interpolation process 
in the low probability regions. 
To avoid interpolation artifacts in the low probability regime, we simply apply a probability threshold to all our 
discrete \logbnpdf{} maps under which all probabilities are set to zero. We chose to set this threshold at the value of 
the 20th percentile of the probability values for each map. 
Typically this threshold corresponds to a probability $< 0.003$, i.e. it only affects the lowest 
probabilities of \bnpdf{} and doesn't vary strongly from model to model. 
\textbf{Varying this threshold did not affect our emulated distributions substantially for values lower than the 
40th percentile of the probability values for each map, as the cut involves the lowest probability regions.}

%%%%%%%%%%%%% FIGURE %%%%%%%%%%%%%
\begin{figure*}
 \plotone{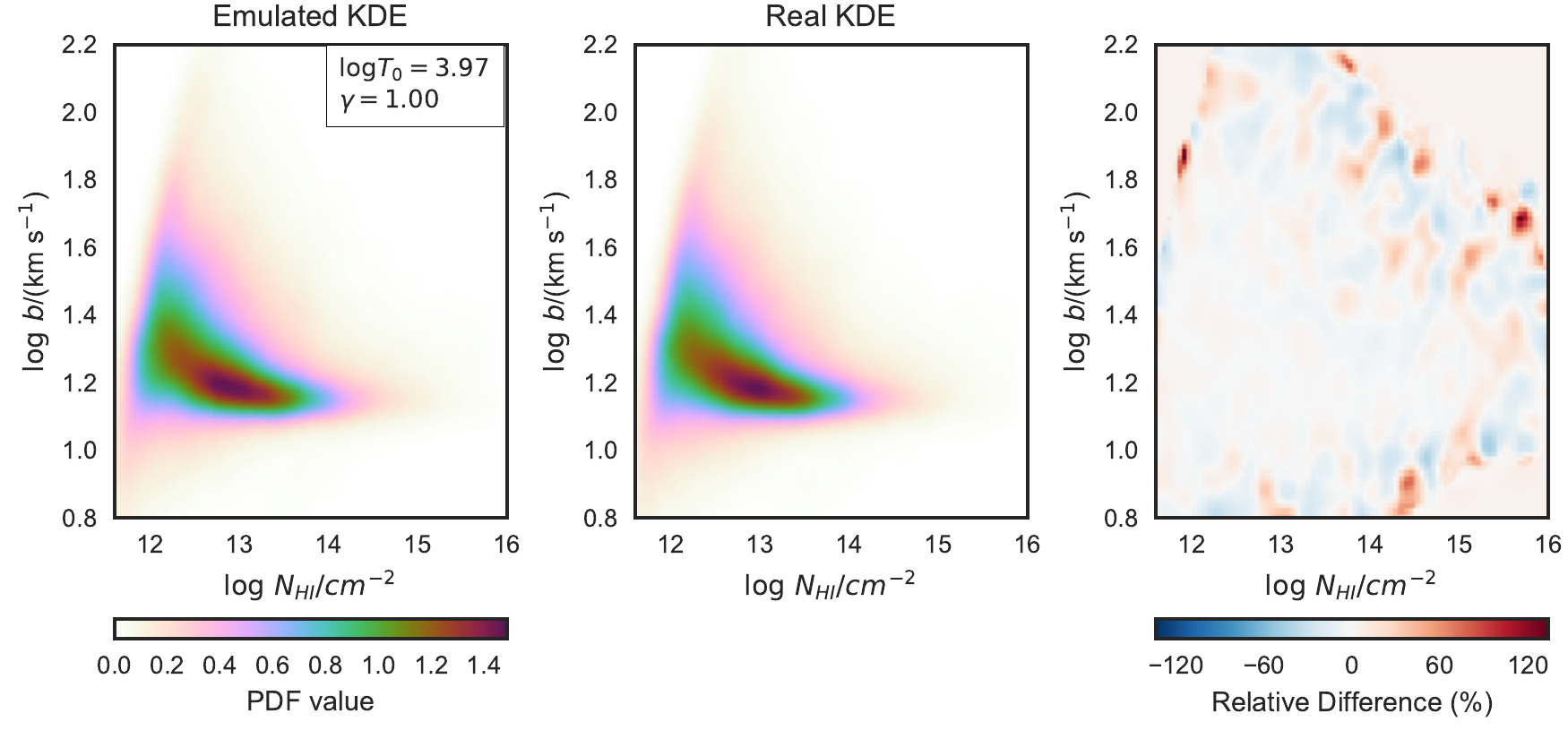}
  \caption{Comparison of interpolated and measured \bnpdf{} for a model in our test grid, i.e. not included in the grid used 
  for constructing the \logbnpdf{} emulator. Thermal parameters are $(\log T_0, \gamma) = (3.972, 1.0)$. 
  \textbf{Left:} \bnpdf{} constructed by interpolating \ac{PCA} components using \ac{GP} interpolation. \textbf{Middle:} 
  \bnpdf{} generated from \ac{KDE} of the \ac{PDF} directly from the \vpfit{} output at the same thermal parameters. 
  \textbf{Right:} The difference of emulated and original \bndist{} relative to the original \bndist{} illustrates that we 
  are able to accurately emulate the \ac{PDF} between our grid points. The fact that we see no relative difference in the edges of the 
  rightmost diagram comes from the fact that we set a density threshold under which the probability was set to zero 
  (see \S~\ref{subsec:pca_decomposition}).}
 \label{fig:int_KDE}
\end{figure*} 

%%% SUBSECTION %%%
\subsection{Emulating the PDF}
\label{subsec:emulator}
Finally, we train a Gaussian process on the \ac{PCA} coefficients for our discrete model grid 
\citep[using \texttt{GEORGE}][]{ambikasaran2016FastDirectMethods}. 
This allows us to generate $\ln$\pbn{} at arbitrary $\log T_0$ and $\gamma$ combinations. 

A Gaussian process is basically a stochastic process for which every finite subset of random variables is normally distributed, i.e. 
it can be fully described by its mean and a covariance function. 
The covariance function is a measure of how much two points in parameter space 
$\vect{\vartheta}_l$ and $\vect{\vartheta}_m$ are covariant, $\vect{\vartheta}$ being a vector with $(\log T_0, \gamma)$ 
in our parameter-space. 
We adopt a standard choice for the covariance $C$, which is a squared-exponential kernel
plus an additional white noise contribution, with the form: 
\begin{equation}
C(\vect{\vartheta}_l, \vect{\vartheta}_m) = \exp \left( -0.5 \, (\vect{\vartheta_l} - \vect{\vartheta_m}) \, C_h^{-1} \, (\vect{\vartheta_l} - \vect{\vartheta_m}) \right) + \sigma_n \delta_\mathrm{lm},
\label{eq:square_exp}
\end{equation} 
where $C_h$ is chosen to be a diagonal matrix with a smoothing length $h_l$ for every dimension, i.e. the characteristic distance beyond
which the covariance between two points drops, and $\sigma_n$ parametrizes the white noise term. 
We chose $h_l$ to be a constant with the value of 20\% of our standard thermal 
grid length in each dimension\footnote{More specifically, prior to the interpolation, our thermal grid was 
renormalized to the range 0 to 1 in each dimension and a kernel size of 0.2 was used. \textbf{See appendix~\ref{app:smooth_length} 
for a motivation of this choice.}} 
(larger than the typical grid separation). 
This guarantees that the interpolation will correlate
coefficients $\Theta_j(T_{0,i},\gamma_i)$ from neighboring points in the grid.

There is an infinite number of functions that satisfy a Gaussian process with a specific mean and covariance, 
but the interpolation (or regression) part comes in once we only select the subset of functions that are constrained to pass 
through a particular set of points. 
In our case, we have a vector of 100 PCA coefficients $\Theta_j(T_{0,i},\gamma_i)$ for each model combination $i$ in our 
grid of 100 simulations. 
Although GP interpolation can be generalized for the case in which the computed \ac{PCA} coefficients 
have uncertainties by having the white noise term $\sigma_n \delta_\mathrm{ij}$ in eqn.~\ref{eq:square_exp}, 
we decided to assume 
that these PCA coefficients have no uncertainty, i.e. we force the interpolation to pass nearly perfectly through the measured 
$\Theta_j(T_{0,i},\gamma_i)$ by setting $\sigma_n$ to nearly zero\footnote{The emulation would not converge when 
setting $\sigma_n=0$, so we adopted the default {\tt TINY} noise value $1.25 \times 10^{-12}$ from the \texttt{GEORGE} library.}. 
This means that our emulator essentially recovers 
the \bndist{} maps perfectly at the thermal grid positions. 

We illustrate the accuracy of our procedure in Figure~\ref{fig:int_KDE}. 
In the left panel we show an emulated \bnpdf{} for a $(\log T_0, \gamma)$ = $(3.972, 1.0)$ combination
between points in our standard grid. The middle panel shows the
true \ac{KDE} based \ac{PDF} from the \vpfit{} output for this thermal model (taken from our test grid). 
The right panel shows the relative difference between the two PDFs,
which scatters around 0 and is typically of the order of 
3\% in probability in the high probability regions, indicating that we can successfully interpolate between models. 
The difference drops to zero in the far edges due to the thresholding of the density described in \S~\ref{subsec:pca_decomposition}. 
There are some peaks in the relative difference close to the edges, that arise simply because the 20th percentile density 
thresholding did not affect the exact same pixels in the real vs. the emulated distribution.

We will further discuss the effect of the \ac{GP} interpolation when performing mock 
measurements in \S~\ref{subsec:inference_test}.

%%% SUBSECTION %%%
\subsection{Parameter Inference}
\label{subsec:likelihood}
We use the \logbnpdf{} emulator as a basis for calculating the likelihood of a dataset given
model parameters. 
The probability of measuring a single absorption line $(N_{\text{HI}, i}, b_i)$ is
given by the \ac{PDF} \bnpdf{}. Thus
the likelihood for measuring a set of $N$ absorption lines ${\log N_{\rm HI},\log b}$ is
\begin{equation}
  \mathscr{L}  = \displaystyle \prod^N_{i} P(\log N_{\text{HI}, i}, \log b_i \given \log T_0,\gamma),
\end{equation}
or in terms of log-likelihood
\begin{equation}
 \ln \mathscr{L} = \sum_i^N\ln P(\log N_{\text{HI}, i}, \log b_i \given \log T_0, \gamma).
 \label{eq:logL}
\end{equation}

Given that our emulator is able to generate model \ac{PDF}s at any
given point within the thermal parameter grid, we simply couple this
log-likelihood to a \ac{MCMC} algorithm to perform Bayesian inference
of the model parameters. For this purpose we use the python package
\texttt{emcee} \citep{foreman-mackey2013emceeMCMChammer} which
implements the affine invariant sampling technique
\citep{goodman2010Ensemblesamplersaffine}. We assumed flat priors for
both parameters which are truncated at the edges of our standard
thermal grid for all \ac{MCMC} runs presented in this paper.

%%%%%%%%%%%%% FIGURE %%%%%%%%%%%%%
\begin{figure}
\plotone{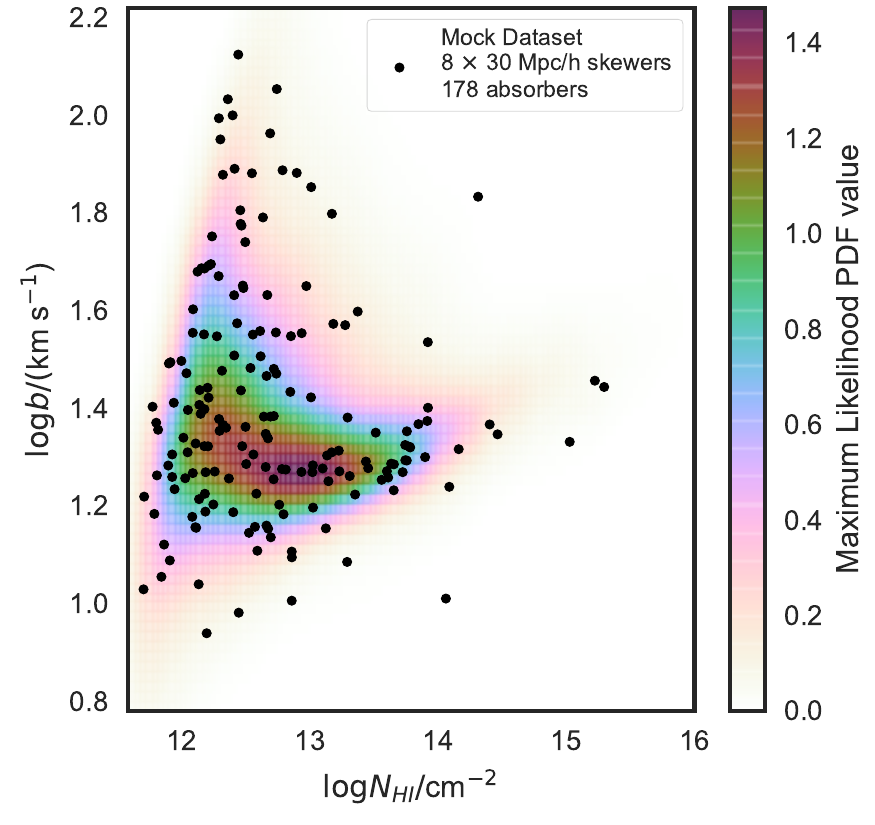}
\caption{A mock realization of a data \bndist{} composed of the absorbers from eight randomly chosen skewers from a simulation with 
$(\log T_0, \gamma) = (4.050, 1.333)$. An emulated \bnpdf{} based on 
the median values of the marginal distributions of the corresponding \ac{MCMC} 
posterior ($\log T_0 = 4.054$ and $\gamma=1.303$, see Figure~\ref{fig:MCMC_corner_interpolated}) is shown for comparison.} 
\label{fig:mockbn}
\end{figure}
The key assumption of the likelihood above is that we treat the \lyaf{} as being an uncorrelated distribution of 
lines such that we can look upon each $\log \NHIt,~\log b$ measurement as a random draw from \bnpdf{}.
We expect that this assumption does not affect our likelihood substantially given the low level 
of spatial correlations in the Ly-a forest \citep{McDonald2006}. We will carry out an inference test in \S~\ref{subsec:inference_test} and asses if this affects mock measurements 
carried out with the PKP method. 

%%%%%%%%%%%%%%%%%%%%%%%%%%%%%%%%%%%
%%%%%%%%%%%%% SECTION %%%%%%%%%%%%%
\section{Testing the Robustness of our Inference}
\label{sec:mock_measurement} 
In this section we test the PKP method by carrying out mock measurements of $\log T_0$ and $\gamma$ 
using \ac{MCMC}. 
First we show one example of a measurement and then we test the robustness of our method 
by examining how the \ac{MCMC} posteriors behave for measurements based on many random realizations of mock datasets 
for the models in our test grid. 

%%%%%%%%%%%%% FIGURE %%%%%%%%%%%%%
\begin{figure}
 \plotone{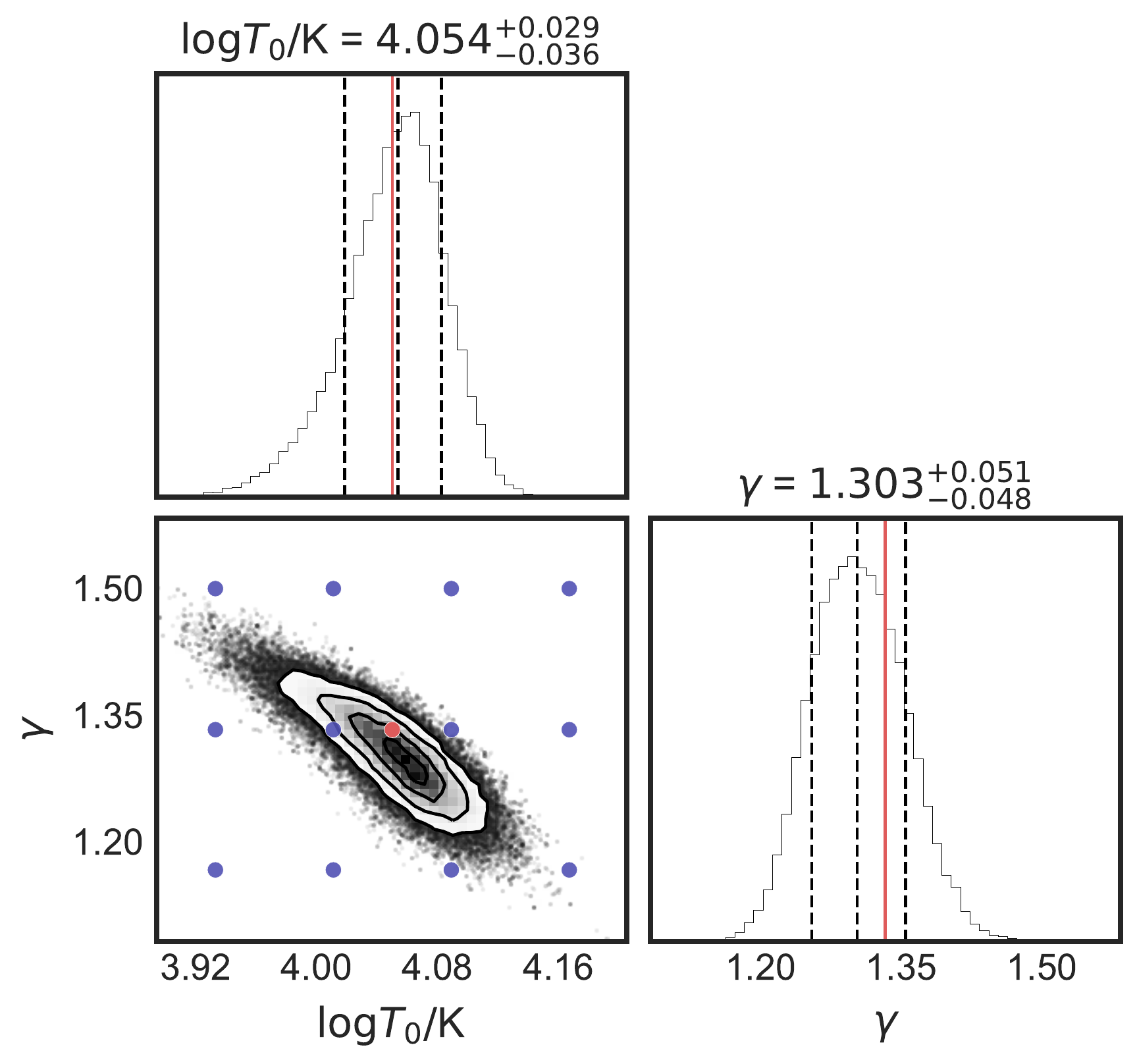}
  \caption{\ac{MCMC} posterior for a mock dataset composed of eight randomly chosen skewers 
  (absorbers shown in Figure~\ref{fig:mockbn}) extracted from a model in our test grid with thermal parameters 
   shown in red. A zoomed in part of the thermal grid used for constructing the emulator on which this measurement is based is shown 
   in blue. The model from which the mock data were chosen (red dot) is not included when constructing the \logbnpdf{} emulator.}
 \label{fig:MCMC_corner_interpolated}
\end{figure}

%%%%%%%%%%%%% FIGURE %%%%%%%%%%%%%
\begin{figure*}
 \plotone{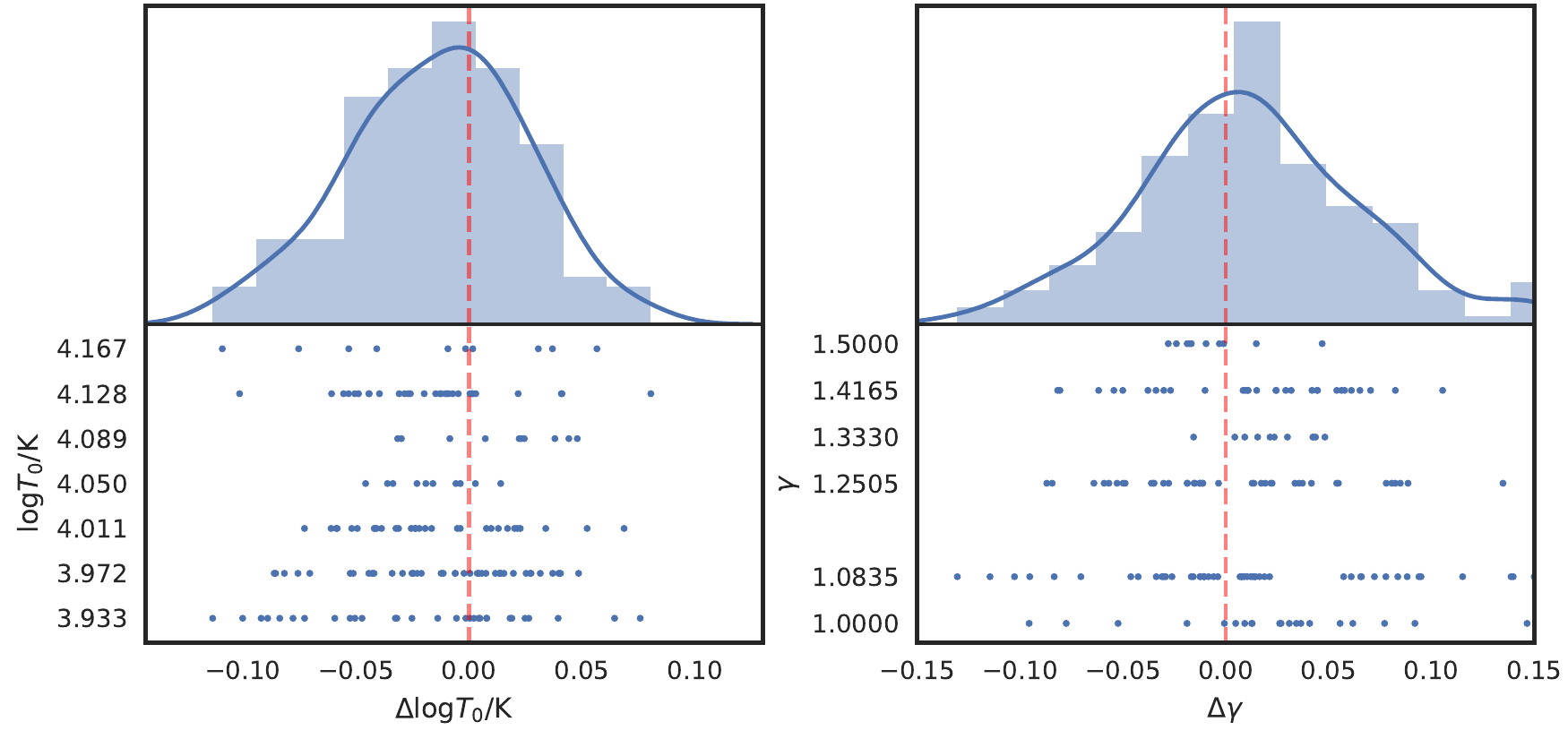}
  \caption{Distribution of the difference between the true values and the medians of \ac{MCMC} posteriors for 10 random realizations 
  of mock datasets with eight skewers each for all the 16 test models (blue in Figure~\ref{fig:thermal_grid}). The likelihoods for these measurements 
  were calculated based on \bnpdf{} generated using our emulator, which did not include these models. The differences between true values and 
  \ac{MCMC} based estimates are shown as blue points in the lower panels for each realization. 
  A histogram of all measurements put together is shown in blue, while the blue line corresponds to a 
  1D KDE of the differences in the histogram. The red dashed line illustrates a perfect measurement.}
 \label{fig:abs_diff_interpolation}
\end{figure*}
%%% SUBSECTION %%%
\subsection{Measurement Example}
\label{subsec:mock_measurement_example}
As an example of a mock measurement we select the absorbers from a sample of eight random 
skewers extracted from a model with $(\log T_0, \gamma) = (4.050, 1.333)$
in our test grid (the blue points in Figure~\ref{fig:thermal_grid}).
The corresponding dataset is shown as black points in Figure~\ref{fig:mockbn}.
For reference, this mock dataset is comparable \textbf{in terms of pathlength to the} redshift range
1.9 to 2.1 provided by a single quasar spectrum in the \citet{Hiss2018} analysis. 
Specifically, this dataset is generated from a pathlength of $240 \Mpc$. While a single 
\lyaf{} at this redshift (between \lya{} and \lyb{} emission peaks) covers $\sim 620 \Mpc$ (from $z = 2.1$ to 1.7). 
In \citet{Hiss2018} the redshift bin used was 1.9 to 2.1, so each quasar spectrum contributed $\sim 295 \Mpc$. 
\textbf{Effectively, due to the masking applied to the data in order to filter possible metal contaminants and 
the pathlength reduction associated with it, our mock dataset corresponds 
to nearly two sightlines in terms of number of absorbers at this redshift range.}

The results of our MCMC inference for this particular mock dataset are shown in Figure~\ref{fig:MCMC_corner_interpolated}. 
\textbf{We observe the well known strong degeneracy in the measurement of $\log T_0$ and $\gamma$, which is a result 
from setting the pivot-point of the \ac{TDR} at mean density} \citep[see e.g. ][]{lidz1, becker1, Walther2018, Hiss2018}. 
We obtain
$\log T_0 = 4.054^{+0.029}_{-0.036}$ and $\gamma=1.303^{+0.051}_{-0.048}$,
whereby the errors are calculated based on the 16th and 84th percentiles of the marginalized distributions 
of the \ac{MCMC} posterior. One observes that this is remarkably close to the true model that the dataset
was drawn from (indicated by the red dot and lines in Figure~\ref{fig:MCMC_corner_interpolated}). 
We can illustrate the inferred model PDF by inputting these measured thermal parameters (i.e. the median
of the individual marginalized posteriors) into into our emulator, retrieving the corresponding \logbnpdf{} 
and computing $\exp($\logbnpdf{}), which is shown by the color coded distribution in Figure~\ref{fig:mockbn}. 

%%% SUBSECTION %%%
\subsection{Inference Test}
\label{subsec:inference_test}
In order to further test the robustness of our method, we perform measurements of $\log T_0$ and $\gamma$ using 10 mock data 
realizations of \bndist s (based on eight random skewers each) 
for each of the 16 models in the test grid. 
Our uncertainties are quantified based on the two dimensional \ac{MCMC} posteriors (see e.g. Figure~\ref{fig:MCMC_corner_interpolated}). 
Testing our measurements by inspecting many realizations of mock datasets will reveal if our method is returning 
valid posterior probability distributions. 

Given that we are dealing with models exactly between our standard grid points, this test will show if interpolation errors
in \logbnpdf{} result in biased measurements. This is a crucial test given that our typical \ac{MCMC} contours have uncertainties 
that are comparable to the characteristic separation between models in our thermal grid, which is illustrated by the
blue grid points shown in Figure~\ref{fig:MCMC_corner_interpolated}. 
Furthermore, an inference test 
will fail, for instance, if our assumption that we can neglect spatial correlations in the Ly$\alpha$ forest in the likelihood
in eqn.~\ref{eq:logL} is incorrect. 

We test if the uncertainties derived from the \ac{MCMC} posteriors are
sensible by carrying out the following exercise. For all of the 160
posteriors, i.e. 16 distinct models times 10 mock realizations of each
model, we quantify how often the true values of the thermal parameters
used land within the 68\% and 95\% confidence regions of the corresponding 2D 
\ac{MCMC} posterior. We observe that the true values are within the
68\% confidence region 68.7\% \textbf{(110/160)} of the time, and that they are within
the 95\% confidence region 96.9\% \textbf{(155/160)} of the time. This convincingly
indicates that our posterior distributions are robust  
and that we are not over or underestimating our
uncertainties. 

As a further test of whether our inference is significantly biased,
we examine the distribution of 
the difference between the true values of $\log T_0$ and $\gamma$ and the median of the marginalized 
distributions of the \ac{MCMC} posteriors: $\Delta \log T_0 = \log T_{0,\text{true}} - \log T_{0,\text{MCMC}}$ and 
$\Delta \gamma = \gamma_{\text{true}} - \gamma_{\text{MCMC}}$. The distributions of these differences are presented in 
Figure~\ref{fig:abs_diff_interpolation}. 
\textbf{We see that the distributions are centered around zero, indicating that any
bias associated with our method is smaller than the resulting uncertainties. Note that in this initial experiment we are 
deliberately only carrying out our tests for the measurement of $T_0$ and $\gamma$, not taking into account 
the correlations with other parameters such as pressure smoothing scale $\lambda_p$ or amplitude of the UVB. 
While certainly important, adding these dimensions to our analysis is beyond the scope of introducing and 
testing our new approach.}

%%%%%%%%%%%%%%%%%%%%%%%%%%%%%%%%%%%
%%%%%%%%%%%%% SECTION %%%%%%%%%%%%%
\section{Pilot Study: A Measurement of Thermal Parameters at \MakeLowercase{z}=2}
\label{sec:actual_measurement} 
The \ac{DM} only models used for our inference test in \S~\ref{subsec:inference_test} use 
an approximation for generating flux skewers which does not capture the full 
physical picture necessary to properly represent the IGM \citep{Sorini2016}. 
\textbf{While \ac{DM} only simulations were sufficient for our initial tests (see \S~\ref{sec:simulations}),} 
for a realistic \textbf{measurement involving real observational data}, one has to use 
hydrodynamical simulations to generate model distributions. 
In this section we apply the PKP method to real \lyaf{} absorption 
line data using a grid of hydrodynamical simulations to model \bnpdf{}.

%%% SUBSECTION %%%
\subsection{The \bnpdf{} from Hydrodynamical Simulations}
\label{subsec:hydro_models}
Following the approach
described in \S~\ref{sec:simulations} and \ref{sec:method}, we now generate models of \bnpdf{} by applying
\vpfit{} to simulated skewers drawn from hydrodynamical simulations of different thermal models.
Hydrodynamic simulations provide the general physical 
conditions that give rise to the \lyaf{} directly from first principles, with exception of reionization effects, 
thus resulting in realistic \bndist s.
Additionally, pressure smoothing of absorbers is accounted for in a 
physical way as opposed to the artificial smoothing of the density field that was used in the DM models. 
The disadvantage associated with hydrodynamical simulations is that, unlike the DM based model,
it is costly to generate large grids in $T_0$ and $\gamma$ at a given redshift, which could pose a problem given the high 
precision our method can achieve. Nevertheless, grids of $\sim 30$ hydrodynamical simulations 
are computationally feasible 
(see below). 

%%%%%%%%%%%%% FIGURE %%%%%%%%%%%%%
\begin{figure}
 \plotone{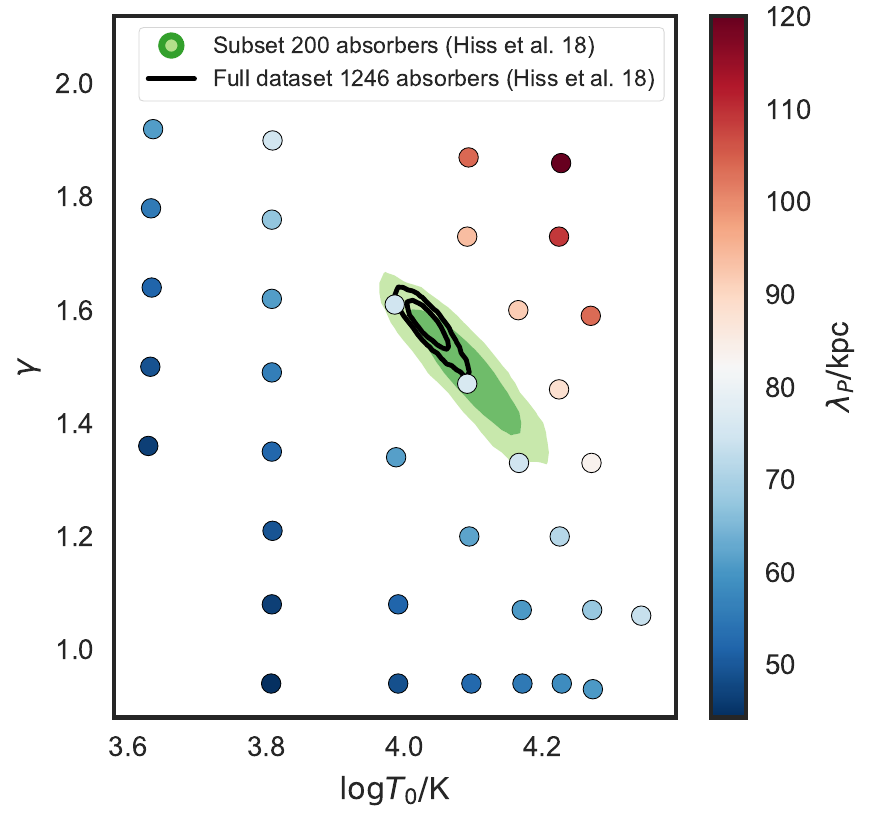}
  \caption{Thermal grid from snapshots of hydrodynamical simulations from the \ac{THERMAL} suite at $z=2$ used in this pilot study. 
  The points are colored based on the pressure smoothing scale $\lambda_P$. For comparison with the characteristic 
  grid separation, we show the measurements 
  (see \S~\ref{subsec:results}) achieved using the full dataset from \citet{Hiss2018} (black contour lines)
  and a subset of 200 absorbers from this dataset (green contours).}
 \label{fig:hydro_grid}
\end{figure}

For the purpose of generating a basis of model \bndist s, we use part of the 
publicly available \ac{THERMAL}\footnote{Url: \url{http://thermal.joseonorbe.com/}} 
suite of Nyx simulations \citep{nyx, Lukic2015} presented in \citet{Hiss2018}. 
The \ac{THERMAL} suite consists of more than 60 Nyx hydrodynamical simulations 
with different thermal histories and $L_{\text{box}} = 20\,\text{Mpc}/h$ 
and $1024^3$ cells based on a \citet{planck} 
cosmology $\Omega_m=0.3192$, $\Omega_{\Lambda}=0.6808$, $\Omega_b =0.04964$, $h=0.6704$, $n_s=0.96$, $\sigma_8=0.826$. 
We chose a grid consisting of a subset of 36 simulation snapshots at $z=2$ with different combinations of 
$T_0$, $\gamma$ and $\lambda_P$ that result from different thermal evolutions \citep{jose1}, shown in Figure~\ref{fig:hydro_grid}. 

Note that, although arbitrary $\lambda_P$ values could be generated in principle, it would require substantial computing power 
to fine-tune the reionization histories to do so. 
As discussed in \citet{Walther2018}, it is difficult to generate physically realistic models without correlating the 
\ac{TDR} parameters and $\lambda_P$, because the pressure smoothing scale depends on the integrated thermal history of the IGM. 
Due to computing time restrictions, we generate only physically motivated $\lambda_P$ which are correlated with the 
\ac{TDR} parameters, i.e. high (low) $T_0$ and $\gamma$ combinations generate large (small) values of $\lambda_P$. 

Following our discussion in \S~\ref{subsec:kde}, we apply the same KDE procedure to the \vpfit{} 
output of our simulations 
and then construct a \logbnpdf{} emulator based on simulated \bndist s (as in \S~\ref{subsec:emulator}).
For the \logbnpdf{} emulation we apply the same \ac{PCA} and \ac{GP} interpolation scheme, adopting smoothing lengths $h$
in the covariance 
(see eqn.~\ref{eq:square_exp}) for the interpolator that is 50\% of the grid size in the 
$\log T_0$ direction and 20\% in $\gamma$ direction. Additionally, for the white noise term in eqn.~\ref{eq:square_exp} 
we chose $\sigma_n = 0.01$, which allows for small deviations \textbf{in the interpolation} at the grid points. 
These changes relative to the DM only emulation 
were arrived at via visual inspection of the emulated PDFs. Specifically, we changed these parameters until no interpolation
artifacts were present throughout the grid. \textbf{A motivation of this choice of white noise contribution is presented in 
the appendix~\ref{app:white_noise}}. 
Additionally, similar to the analysis of mock datasets in
\S~\ref{subsec:inference_test}, 
we checked if we accurately recover the thermal parameters at the grid positions and 
found that the results were unbiased.
This indicates that the different Gaussian process smoothing parameters and white noise term added when using hydrodynamic
simulations do not significantly bias our inference. 

%%% SUBSECTION %%%
\subsection{Absorption Line Dataset}
\label{subsec:dataset}
In order to carry out a measurement, we use the absorption line data from \citet{Hiss2018} 
which consists of 1246 absorption lines\footnote{In line with our approach in \S~\ref{subsec:forward_modeling}, 
\citet{Hiss2018} excluded absorbers that have relative uncertainties worse than 50\% in $b$ or \NHI{} 
from their observational dataset. For consistency, the same recipe was applied to the lines of sight 
extracted from our hydrodynamical simulations.} at $1.9 \leq z < 2.1$. 

One problem that could bias the results of our method are outliers with low \bvel{} in the \bndist{}. 
\citet{Hiss2018} argued that these are narrow lines added by \vpfit{} in 
order to decrease the $\chi^2$ of the fit in blended absorption features, and 
unidentified metal absorbers wrongly assumed to be \lya{} lines (as observed by \citet{schaye3, rudie1}). 
Blending artifacts should not have a severe impact on our measurements, 
as a proper forward modeling of the simulated sightlines should include the same sort of contamination in our model \bnpdf{}. 

As for dealing with metal line 
contamination, the dataset used was carefully masked for metal absorption systems, as described in \citet{Hiss2018, Walther2018a}. 
The severity of metal line contamination is strongly redshift dependent, 
as the identification of metal absorbers in the \lyaf{} becomes increasingly difficult at higher redshift (and nearly unfeasible at 
$z \gtrsim 3.5$) due to line blanketing as the effective optical depth of the \lya{} forest
increases. \textbf{In our case, the contamination should be relatively mild, given that metal line absorbers are 
more easily identified at lower redshifts and that these data were previously masked for potential contaminants using different automatic 
and interactive techniques}. Nevertheless there are remaining unidentified contaminants, 
that have to be excluded with some sort of outlier 
rejection. 

\begin{figure}
 \plotone{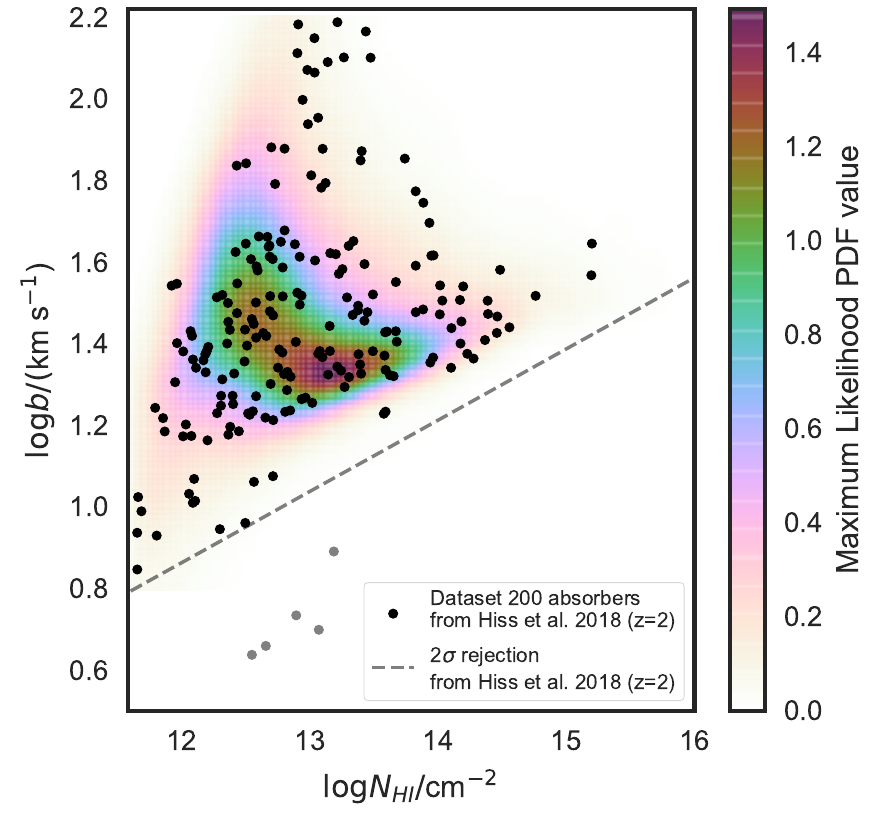}
  \caption{A subset of the \bndist{} from \citet{Hiss2018} composed of 200 randomly chosen absorbers (black points). 
  To avoid possible narrow line contaminants (gray points) only absorbers with $b$ above the extrapolated 2$\sigma$ 
  rejection line from \citet{Hiss2018} were chosen (gray dashed line). An emulated \bnpdf{} based on 
 the median values of the marginal distributions of the corresponding \ac{MCMC} 
 posterior is shown for comparison.}
 \label{fig:hydro_fit}
\end{figure}

In \citet{Hiss2018} they implemented an iterative 2$\sigma$ rejection procedure based on \citet{rudie1} that rejects potential 
narrow line contaminants in the range $12.5<\log \NHIt/\text{\cmtwo}<14.5$. 
For simplicity, we decided to extrapolate the 2$\sigma$ rejection line defined in \citet{Hiss2018} 
to the region $11.5<\log \NHIt /\text{\cmtwo}<16$ (shown as a gray dashed line in Figure~\ref{fig:hydro_fit}) 
and discard all absorbers with $\log b$ lower than this line. 
Alternatively, one could implement a more elegant 
outlier modeling method such as the one used by \citet{Telikova2018}, but here we opt for this simpler approach. 

%%%%%%%%%%%%% FIGURE %%%%%%%%%%%%%

The dataset from \citet{Hiss2018} has a size of 1246
absorbers, and we have intuition from \S~\ref{subsec:inference_test}
that this dataset size would result in percent level precision, i.e. smaller 
than the spacing between our thermal grid points\textbf{, making our inference susceptible to 
interpolation uncertainties}. 
We thus decided to randomly choose a set of 200 absorbers from this dataset, hence with a similar number of
lines as the mock dataset of eight skewers described in
\S~\ref{sec:mock_measurement}. 
In contrast to the SNR modeling done in \S~\ref{subsec:forward_modeling} using a constant value 
of 63 per 6\,\kms{}, 
we randomly chose the SNR from the real sightlines for the mock spectra from hydrodynamical simulations 
to better represent the noise distribution within the data (exactly as was done in \citealt{Hiss2018}). 
Because of this approach, it makes more
sense to chose a random subset of absorbers rather than selecting a random subset of quasar sightlines. 
\textbf{For a discussion about how our results differ if we randomly choose QSO sightlines instead of absorbers 
please refer to the appendix~\ref{app:subsampling}}.

To understand how our uncertainties compare to the typical separations between
points in our thermal parameter grid we show two sets of $\log T_0$-$\gamma$ measurements in
Figure~\ref{fig:hydro_grid}. We will explain in detail how these
contours were measured in the next section. But for the sake
of the current discussion, note that the green contours
result from analyzing a dataset of 200 absorbers, resulting 
a precision comparable to our characteristic grid separation; whereas, 
the black contours show a measurement using the complete dataset of 1246
absorbers. Clearly, using the full dataset results in an uncertainty 
substantially smaller than our grid spacing, which
indicates that interpolation errors could be a significant issue. 
Given the exquisite precision delivered by the PKP method and
the size of existing datasets, it is challenging to generate a grid of hydro simulations
fine enough to do justice to the implied precision. Nevertheless, we believe that
this is computationally within reach and will enable measurements
of the thermal state of the IGM with unprecedented precision. 

%%% SUBSECTION %%%
\subsection{Results}
\label{subsec:results}
In order to measure $\log T_0$ and $\gamma$, we carry out the same Bayesian measurement as described in \S~\ref{sec:mock_measurement}, 
this time using real data combined with \pbn{} emulated from hydrodynamical simulations. 
The subset of 200 absorbers from \citet{Hiss2018} are shown as black points in Figure~\ref{fig:hydro_fit},
whereas the five gray points are the corresponding fraction of absorbers 
that are rejected. 
The green contours in Figure~\ref{fig:hydro_grid} shows the MCMC posterior resulting from analyzing these
data, from which we measure $\log T_0 = 4.092^{+0.050}_{-0.055}$ and $\gamma=1.49^{+0.073}_{-0.074}$,
whereby the errors are calculated based on the 16th and 84th percentiles of the marginalized distributions. 
\textbf{We explore how this inference behaves for 
different random realizations of 200 absorbers in the appendix~\ref{app:subsampling}.}
As before, we emulate the \bnpdf{} at these measured values which is shown as the color coded
distribution in Figure~\ref{fig:hydro_fit}. 

\textbf{Additionally, we} carried out the same measurement using the full dataset of 1264 lines from \citet{Hiss2018}.
As discussed in \S~\ref{subsec:dataset}, due to the current separations in our model grid, we
have concerns about interpolation error at such a high level of precision.
Nevertheless, we wanted to illustrate the kind of precision achievable using existing data.
With these caveats, we measure $\log T_0 = 4.034^{+0.022}_{-0.019}$ and $\gamma=1.576^{+0.026}_{-0.032}$.
The corresponding contours are shown in black in 
Figure~\ref{fig:hydro_grid}. 
Importantly, compared to the measurement using a subset of these data, the uncertainties are smaller by a 
factor of approximately $\sqrt{6}$, which is the expected scaling due to the relative sizes of the datasets. 

% subsubsection
\subsubsection{Comparison with Cutoff Fitting Results}
These PKP based results can be compared to the \citet{Hiss2018} measurements from the
same dataset using the cutoff fitting approach. 
From the marginalized distributions of the \citet{Hiss2018} Monte Carlo based posteriors, 
they measured $\log T_0 = 4.137^{+0.050}_{-0.074}$ and $\gamma=1.47^{+0.12}_{-0.10}$ at $z=2$ using
the 1264 \lya{} absorbers. 

%%%%%%%%%%%%% FIGURE %%%%%%%%%%%%%
\begin{figure}
 \plotone{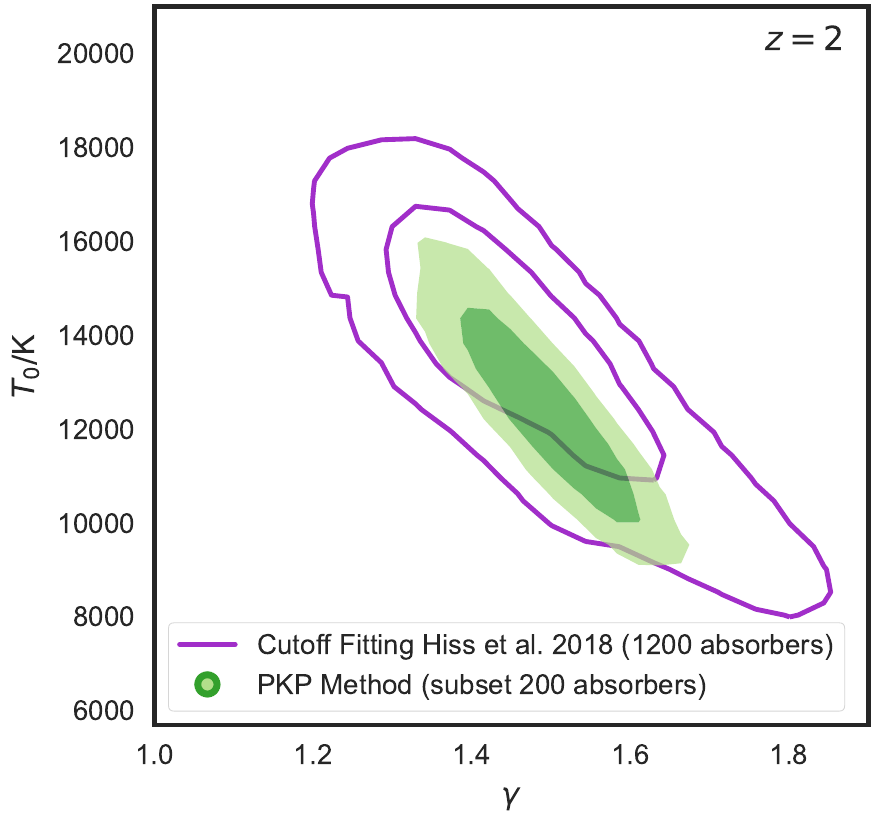}
  \caption{Comparison of the thermal parameter constraints from \citet{Hiss2018} using the cutoff fitting method 
  (purple contour lines), and 
  our measurement using the PKP method (green contours). While the original dataset from \citet{Hiss2018} has a size of $1246$ absorbers, only 
  845 are actually used for cutoff fitting due to the fact that only absorbers with 
  $12.5 \leq \log(\NHIt{}/\text{\cmtwo{}}) \leq 14.5$ 
  and $8 < b/\kms < 100$ are 
  used. The cutoff fitting results are shown as purple contour lines. When using the PKP method described in this study, we 
  achieve higher precision (green contours) while using a random subset of 200 absorbers from their data.}
 \label{fig:cutoff_comparison}
\end{figure}

As stated in \S~\ref{subsec:results}, when applying our new method to a subset of 200 absorbers from their dataset, 
we measure 
$\log T_0 = 4.092^{+0.050}_{-0.055}$ and $\gamma=1.49^{+0.073}_{-0.074}$. 
In Figure~\ref{fig:cutoff_comparison} we compare our PKP based measurement using just 200
absorbers from \citet{Hiss2018} (green shaded contours) 
to the cutoff fitting measurement from \citet{Hiss2018} using the full dataset 
(purple contours). 

A direct comparison of these measurements based on the size of the dataset used is challenging, because 
both methods use different cuts in the data. While we use all absorbers within the allowed fitting range, 
the cutoff fitting method only uses the absorbers within 
$12.5 \leq \log(\NHIt{}/\text{\cmtwo{}}) \leq 14.5$ 
and $8 < b/\kms < 100 $. 
In \citet{Hiss2018} this reduces the initial dataset of 1264 to 845 absorbers which are 
effectively used for cutoff fitting. 

As described in \S~\ref{subsec:results}, using the complete dataset
results in a dramatic improvement in the precision compared to
\citet{Hiss2018}\footnote{This comparison may seem unfair since
\citet{Hiss2018} marginalized their results over different pressure
smoothing scales $\lambda_P$, which we do not do in this
work. Nevertheless this marginalization did not significantly impact
their measurement precision, i.e. their uncertainties in $T_0$ and
$\gamma$ are dominated by the statistical error on the cutoff
parameters.}. 
This improvement comes from the fact that the constraining power of
the cutoff method depends only weakly on the number of absorbers in
the \bndist{}, as discussed in detail by \citet{schaye3} (see their
Figure 14), and hence its precision does not scale as $\sqrt{N}$ as
one would naively expect. 
In contrast, the advantage of the PKP method is that it delivers a precision which scales
approximately as $\sqrt{N}$, delivering higher precision for larger datasets. 

For a more direct comparison one can calculate what uncertainties we would expect for a dataset 
of 845 absorbers, i.e. the exact number of absorbers effectively used for cutoff fitting. 
Under the assumption of $\sqrt{N}$ scaling, our representative uncertainties for a dataset of 200 absorbers, 
for example $\sigma_{\log T_0} = 0.055$ and $\sigma_{\gamma}=0.074$, become smaller by a factor $\sqrt{845/200}$, i.e 
$\sigma_{\log T_0} = 0.027$ and $\sigma_{\gamma}=0.036$. In this case our result would be around factor of two 
in $\log T_0$ and a factor of nearly three in $\gamma$ more precise 
than cutoff fitting for 845 absorbers. 

Indeed, the main limitation in PKP precision, which we have already encountered for the
current dataset, is the number of simulations required to generate a model grid dense enough to deliver the implied
precision. However, we believe this is a surmountable problem given currently available computational resources. 

Finally, we note that another complication associated with the cutoff 
fitting method is that one has to adopt a value of 
the column density $N_{\text{HI},0}$ that corresponds to the mean density in order to relate the 
minimal Doppler parameter at this density $b_0 = b_{min}(N_{\text{HI},0})$ to $T_0$. With this new approach 
we circumvent this issue, as we are sensitive to the shape of the \bndist{} at all column 
densities. 
Furthermore, \citet{Hiss2018} showed that cutoff fitting is sensitive to the details of the
iterative cutoff fitting method (least squares or mean deviation minimization), which can lead to differences in the results.
In contrast, the Bayesian likelihood (eqn.~\ref{eq:logL}) that provides the underpinnings of PKP does
not require that one make these somewhat arbitrary choices.

%%%%%%%%%%%%%%%%%%%%%%%%%%%%%%%%%%%
%%%%%%%%%%%%% SECTION %%%%%%%%%%%%%
\section{Discussion and Summary}
\label{sec:summary}
In this work we introduced a new method for inferring thermal parameters from the 
\bndist{} of \lyaf{} absorbers in the IGM, the PKP method. In contrast to a large body of previous work focused on
analyzing a small subset of lines to fit the lower cutoff of the \bndist{}, our new approach utilizes all available data and exploits
parameter sensitivity encoded in the full shape of this distribution. We generated a large grid of simulations of
the \lyaf{} encompassing a range of different thermal parameter models, and fit the resulting mock spectra with
\vpfit{}, generating a large database of absorption lines for each model. Our new method applies KDE to
sets of discrete absorption lines to generate model \bndist{} PDFs, then uses a PCA decomposition to create
an emulator for this distribution which can be evaluated at any location in thermal parameter space. Using this
emulator, we introduced a Bayesian likelihood formalism enabling parameter inference via MCMC. We conducted a
pilot study demonstrating the efficacy of this new approach \textbf{in the limit of a two dimensional $T_0$ and $\gamma$ measurement,} 
whereby real observational data at $z=2$ was compared
to a grid of hydrodynamical simulations.
The primary results of this work are: 

\begin{enumerate}
\item Using 160 mock measurements we demonstrated that our statistical inference procedure delivers unbiased 
estimates of thermal parameters and reports valid uncertainties.

\item Our new method was applied to real observational data to measure the parameters of the \ac{TDR} at $z=2$.
We found $\log T_0 = 4.092^{+0.050}_{-0.055}$ and $\gamma=1.49^{+0.073}_{-0.074}$ using just a subset 200 absorbers 
from the dataset of \citet{Hiss2018}, which roughly corresponds, \textbf{in terms of pathlength, to} a single \lyaf{}
spectrum at $z\simeq 2$. 

\item For current dataset sizes at $z$=$2$, the PKP method can already
deliver a precision on $\log T_0$ ($\gamma$) nearly two (three)
times higher than the cutoff fitting method.
\end{enumerate}

In the future this method could be expanded to include other
parameters that affect the shape of the \bndist{}. One could model
different thermal histories by including the pressure scale
$\lambda_P$ as a free parameter, allow the mean flux $\bar{F}$ to
vary, which would constrain the \ac{UVB}, or analyze IGM models with
additional physics such as blazar heating \citep{puchwein1, Sironi2014, lamberts1} or galaxy 
formation feedback \citep{Sorini2018}. 
Our new methodology is readily applicable to the $z > 2$ Ly$\alpha$ forest, as shown by our pilot study at $z=2$,
as well as to existing Hubble Space Telescope Cosmic Origins Spectrograph 
(HST/COS) \ac{UV} spectra \citep[e.g.][]{Danforth2013, Danforth2016} that probes the \lyaf{} at 
$z\lesssim 0.5$. Indeed, measuring the thermal state of the IGM at these low redshifts with high precision
could help clarify the nature of the discrepancy of
the \bndist{} between observations and hydrodynamical simulations that have been recently highlighted \citep{Viel2017, Gaikwad2017, Nasir2017}. 

%%%%%%%%%%%%%%%%%%%%%%%%%%%%%%%%%%%
%%%%%%%%%% Acknowledgments %%%%%%%
\acknowledgments We thank the members of the ENIGMA group at MPIA and
UCSB for helpful comments on an early version of this manuscript. HH
also acknowledges the members of office 217 for the fruitful
discussions. Special thanks also to Martin White for providing the
collisionless DM simulations used in this work. Finally we
acknowledge Stu Mackenzie for inspiring our choice of colormaps.
Calculations presented in this paper used the hydra and draco clusters
of the Max Planck Computing and Data Facility (MPCDF, formerly known
as RZG). MPCDF is a competence center of the Max Planck Society
located in Garching (Germany).
Some data presented in this work were obtained from the Keck
Observatory Database of Ionized Absorbers toward QSOs (KODIAQ), which
was funded through NASA ADAP grants NNX10AE84G and NNX16AF52G along
with NSF award number 1516777. 
Some of the measurements use
observations collected at the European Southern Observatory.
This research used resources of the
National Energy Research Scientific Computing Center (NERSC), which is
supported by the Office of Science of the U.S. Department of Energy
under Contract no. DE-AC02-05CH11231.

% BIBLIOGRAPHY
\bibliography{references.bib}

\appendix
%%%%%%%%%%%%%%%%%%%%%%%%%%%%%%%%%%%
%%%%%%%%%%%%% SECTION %%%%%%%%%%%%%
\section{Thermal Sensitivity Animations}
\label{app:animations}

%%%%%%%%%%%%% FIGURE %%%%%%%%%%%%%
\begin{figure}[h!]
  \plottwo{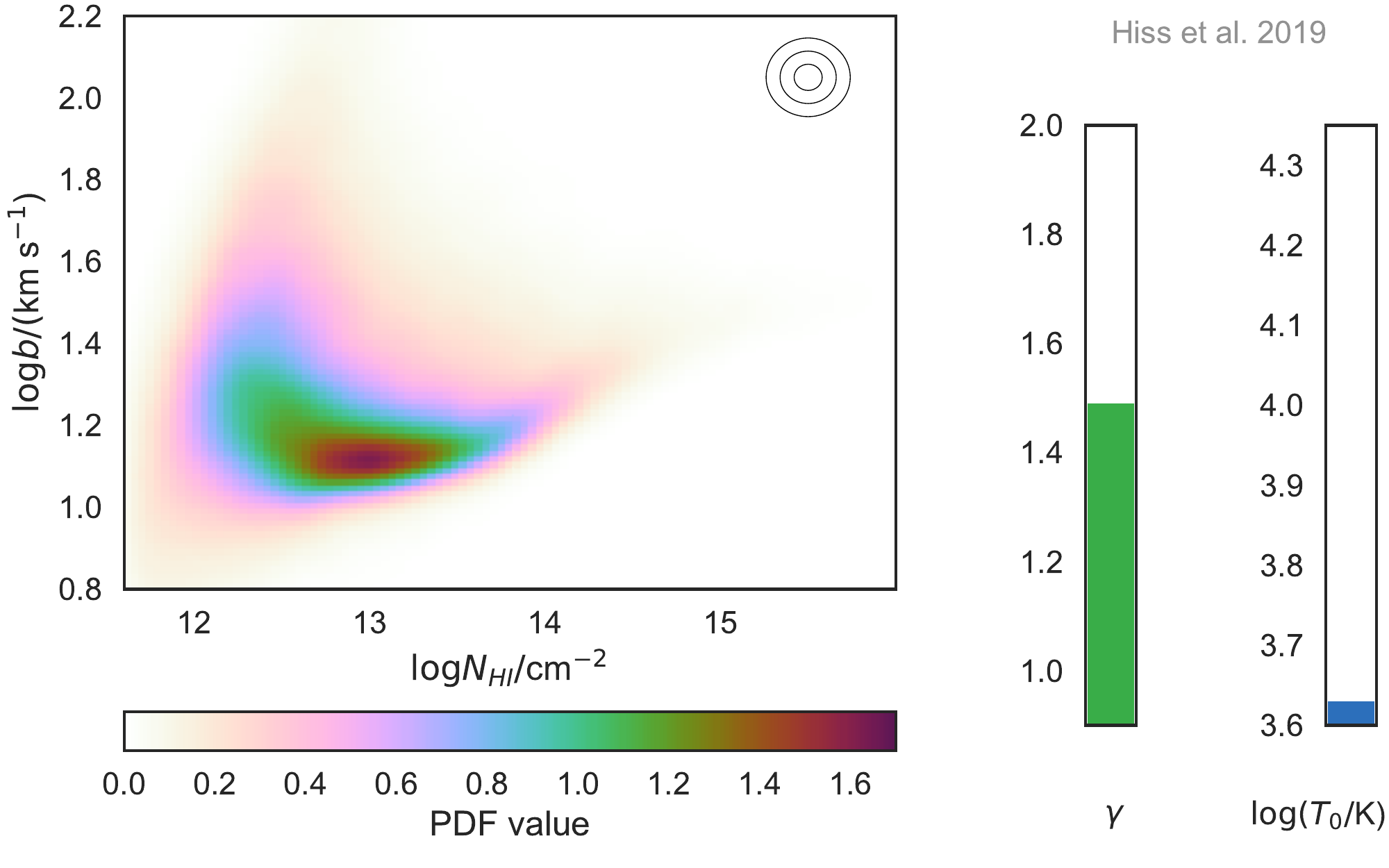}{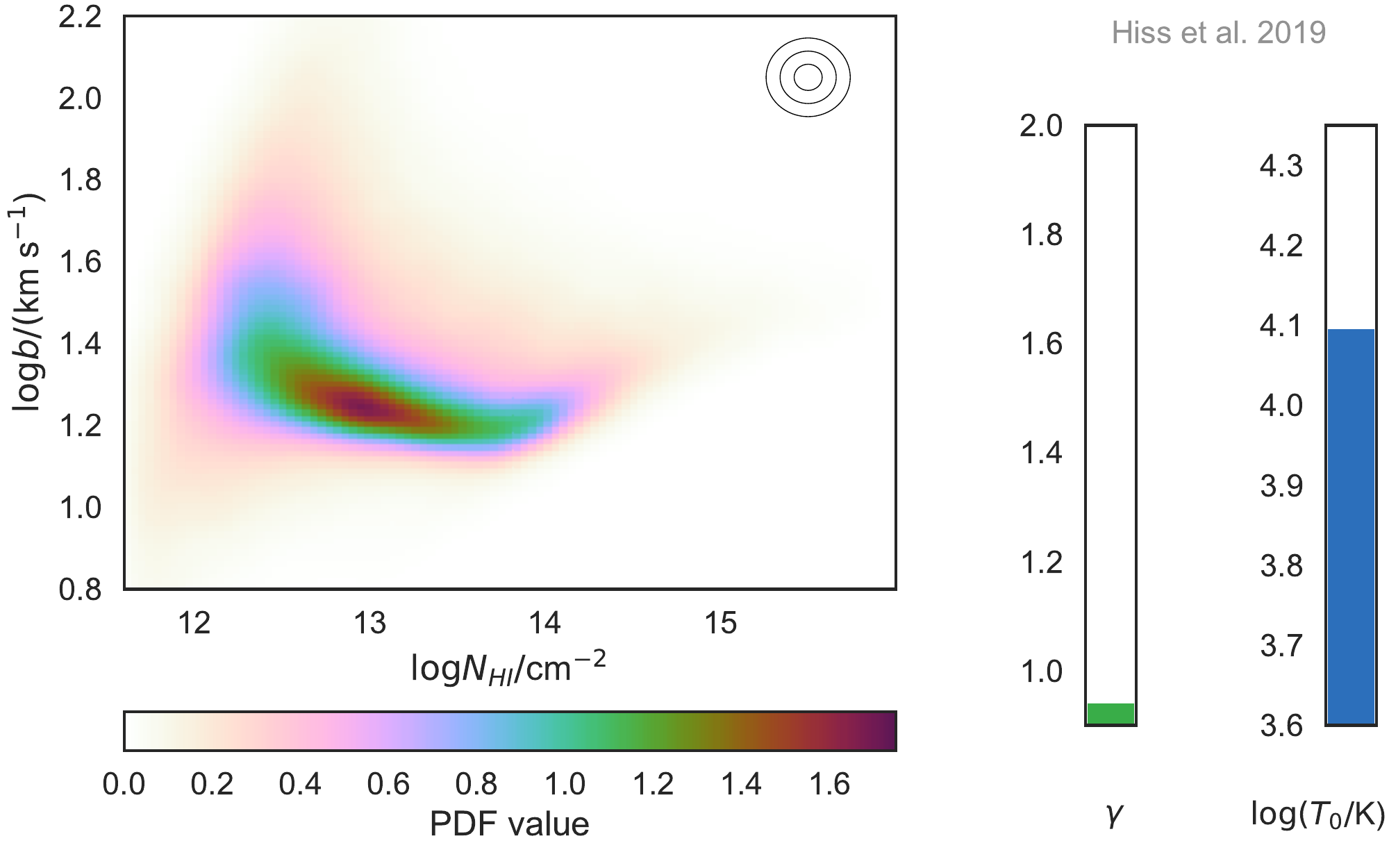}
  \caption{These Figures are meant to be viewed as animations in the HTML version of this manuscript (available in the refereed version only). Both animations were generated 
  using our emulator based on hydrodynamical simulations described in \S~\ref{subsec:hydro_models}. 
  \textbf{Left:} Change of the shape 
  of the \bndist{} when changing $\log T_0$ from 3.63 to 4.26 in ten equal steps at a fixed $\gamma=1.49$. Similar to the 
  effect illustrated in the upper panels of Figure~\ref{fig:thermal_sensitivity}, increasing $\log T_0$ 
  shifts the distribution toward higher $b$. 
  \textbf{Right:} Change of the shape 
  of the \bndist{} when changing $\gamma$ from 0.94, 1.87 in ten equal steps at a fixed $\log T_0=4.09$. 
  Similar to the 
  effect illustrated in the lower panels of Figure~\ref{fig:thermal_sensitivity}, increasing $\gamma$ 
  mainly tilts the distribution at $\log (\NHIt$/\cmtwo{})$>13$. For both panels, the concentric 
  rings on the top right represents the KDE bandwidth chosen (1, 2, and 3$\sigma$). 
  Note that there is a change in the pressure smoothing scale included in these animations which is not 
  explicitly shown. As shown in Figure~\ref{fig:hydro_grid}, this particular emulator was built based 
  on a grid that correlates the thermal parameters and the pressure smoothing scale due to the dependence of $\lambda_P$ 
  on the integrated thermal history of the IGM. Broadening across this characteristic length is responsible for the turn 
  over in the distribution visible at $\log (\NHIt$/\cmtwo{})$<13$ when increasing either $\log T_0$ or $\gamma$.}
 \label{fig:animationT0}
\end{figure}

\bfseries{
\section{Choice of Emulation Hyperparameters}
\label{app:hyperparams}
\subsection{Emulator Smoothing Length $h_l$}
\label{app:smooth_length}
To motivate the choice of $h_l=0.2$ for our emulator smoothing length for the \ac{DM} only models, we compare the true PDF for a model with $\log T_0 = 4.128$ and $\gamma = 1.4165$ to the emulated PDF at the same thermal parameters using different smoothing lengths in Figure~\ref{fig:smoothing_length}. 
This particular model is not included in the emulator building process, but is part of our test grid (see Figure~\ref{fig:thermal_grid}), and was chosen to lie as far away from grid points as possible. 
We show the true PDF, i.e. the one computed directly from the \bndist{}, in the upper left panel and the difference between emulated and real PDF for different $h_l$ in the other panels. 
Essentially, the emulated PDF differs substantially from the true one when choosing very small (smaller than grid separation, i.e. 
emulator does not correlate neighboring models) 
and very large $h_l$ ($>$factor of 3 grid separation). The emulator shows a stable performance in the 
intermediate range $0.03 \leq h_l\leq 0.3$ which implies that the choice of $h_l=0.2$ is adequate. 
Note that this example shows the worst case scenario where the emulated PDF is the farthest away from the grid points and 
the interpolation has the highest uncertainty. 

\begin{figure}
\centering
\plotone{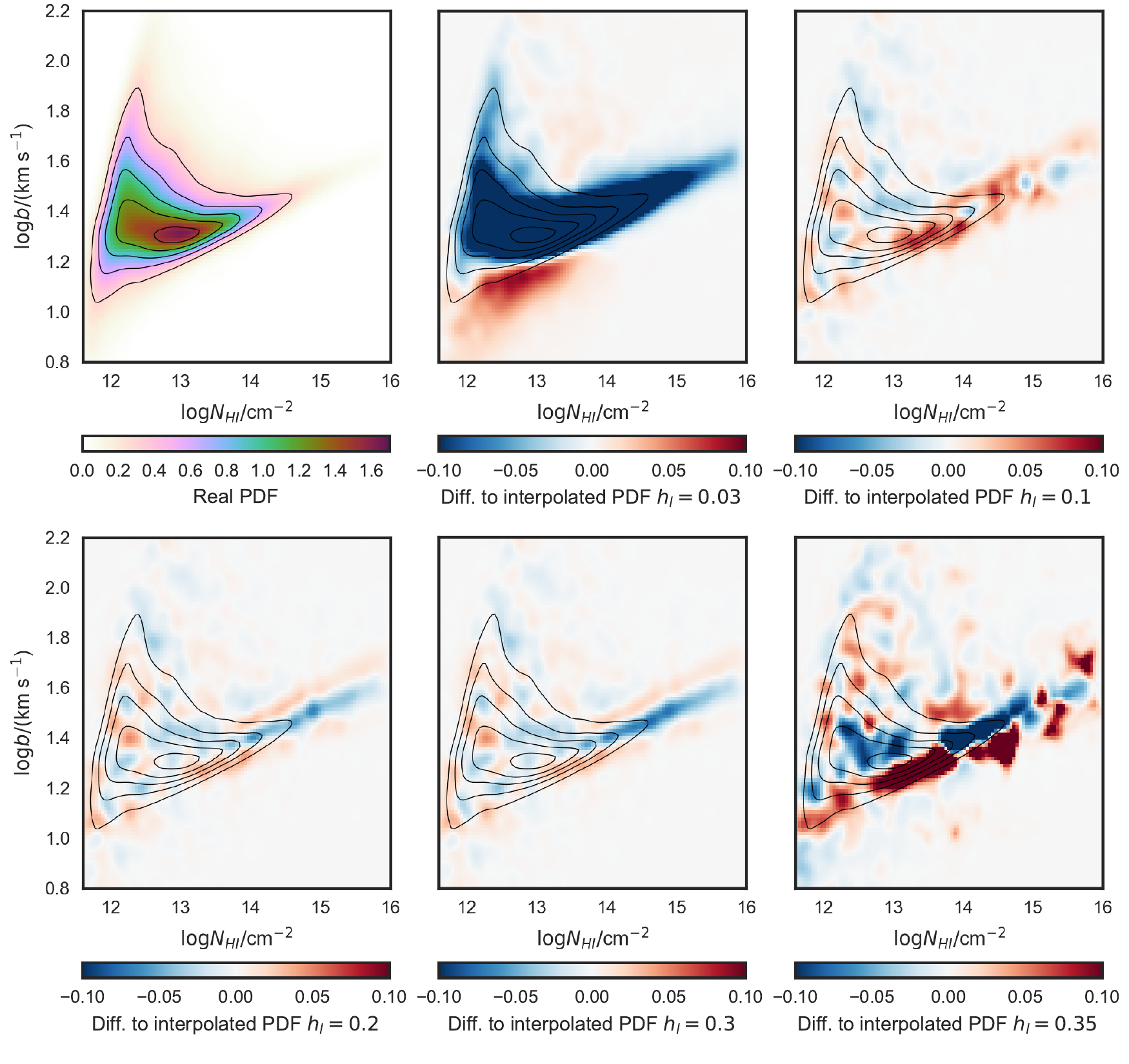}
\caption{The difference between true and emulated maps for a model with $\log T_0 = 4.128$ and $\gamma = 1.4165$ (not included in the \ac{DM} only emulation grid and maximally far away from points in the grid). The emulated PDFs were constructed from emulators using different smoothing lengths. 
Using a smoothing length that is too small results in an interpolation that does not take into account close grid points, while a 
large smoothing length introduces artifacts. We observe small fluctuations in comparison with the true PDF 
for intermediate $0.03<h_l<0.3$ for our \ac{DM} only emulation scheme.}
\label{fig:smoothing_length}
\end{figure}

\subsection{White Noise Contribution $\sigma_n$}
\label{app:white_noise}
In \S~\ref{subsec:hydro_models} we state that we chose the value $\sigma_n=0.01$ for the hydrodynamic simulation grid based on visual inspection, because we observed clear interpolation artifacts in a few places inside the thermal grid when adopting no white noise contribution. To explore the effect of this choice, we show in Figure~\ref{fig:white_noise} one example with $\log T_0=3.9$ and $\gamma=1.19$ that generated such artifacts. The upper left panel shows the color coded map and contours 
for the choice used in this study. All other panels represent different choices of white noise contribution. 
Note that this choice of $\log T_0$ and $\gamma$ represents the worst case of artifacts we encountered within the grid, 
and it corresponds to a location where the interpolation covers a substantial gap in parameter space. 
Unfortunately, we do not have the option of generating extra models as we did with the DM only simulations, i.e. the true PDF 
at this grid position is unknown, but Figure \ref{fig:white_noise} indicates that (in this worst case scenario) 
the general shape of the emulated \bndist{} does not present artifacts for $\sigma_n>0.005$ and keeps its general shape 
until $\sigma_n$ is large ($>0.1$) and the interpolation has so much freedom in the grid points that the shape 
of the \bndist{} loses information about the thermal state of the gas. 

\begin{figure}
\centering
\plotone{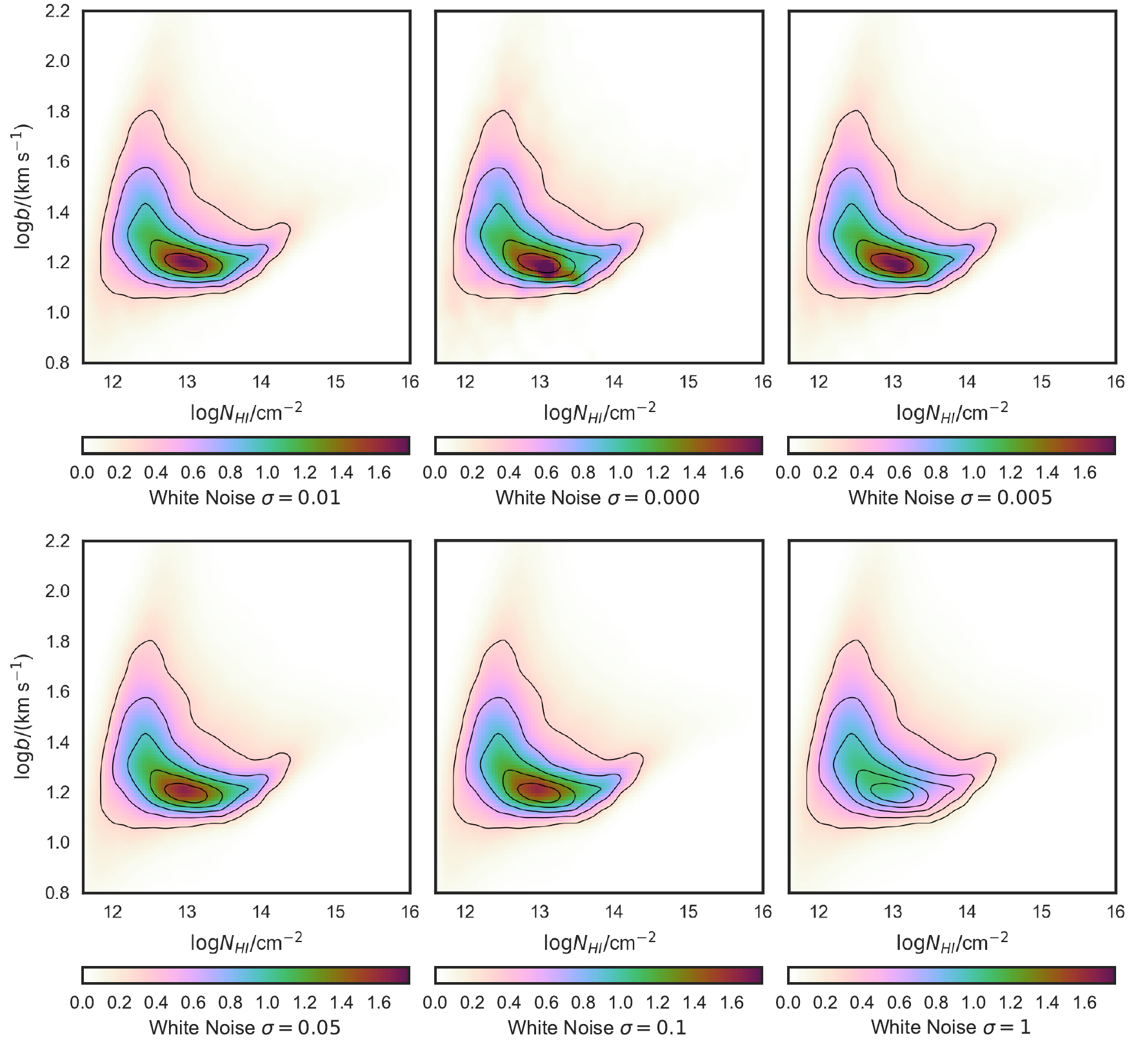}
\caption{Emulated \bndist{} for $\log T_0 = 3.9$ and $\gamma = 1.19$ using the hydrodynamic grid for different values 
of the white noise term $\sigma_n$. The emulated PDF resulting from an emulation using our fiducial choice of $\sigma_n=0.01$ is shown in the upper left panel as a color coded map and corresponding contours. All other panels show the \bndist{}, but emulated using different white noise contributions. The contours of our fiducial choice are shown for comparison in all panels. This figure illustrates 
that allowing no freedom for the interpolation at the grid points results in interpolation artifacts in this particular position 
(between grid points). Additionally, allowing too much freedom results in loss of information about the thermal state.}
\label{fig:white_noise}
\end{figure}

\section{Effect of Different Data Subsampling Methods}
\label{app:subsampling}
As stated in \S~\ref{subsec:dataset}, we chose to draw 200 absorbers randomly from the dataset of \citet{Hiss2018}, because the models used to construct the \bndist{} PDFs have a mixed SNR with a distribution based on our data. 
This approach could pose a problem, as random picking across the full dataset essentially removes correlations between absorbers 
in the same spectra. 
We showed in section \ref{subsec:inference_test}, using \ac{DM} only simulations, that our inference is robust in the case of a fixed \ac{SNR} and mock datasets composed of 8 randomly drawn skewers, i.e. correlations are included and the \ac{SNR} 
does not affect our inference test. 
To understand if these effects play a role in the measurement presented in \S~\ref{subsec:results}, one should investigate the effects of picking random QSO sightlines instead of random absorbers, given that our likelihood is agnostic to correlations 
between absorption lines. 
In the following paragraphs we will explore both approaches. 

To test if our inference is influenced by randomly choosing absorbers, we generated another 200 realizations of 200 randomly chosen 
absorbers from the full dataset and carried out the same inference as in \S~\ref{subsec:mock_measurement_example}. 
Note that, while the same absorbers are present in different realizations, absorbers are picked 
without replacement  such that the same absorbers does not appear more than once in each individual realization. 
As a measure for how consistent the measurements of all these 
realizations are with each other, given that we do not know the true value, we compare the measurements of each realization to the measurement using the full dataset presented in \S~\ref{subsec:results}. 
We observe that the measurements from the full dataset ($\log T_0 = 4.034$ and $\gamma=1.576$) are within the 1$\sigma$ contour of the 2D posteriors of these realizations 65\% (129/200) of the time,
and within the 2$\sigma$ contour of the 2D posterior 96\% (192/200) 
of the time. 
This implies that our inference is consistent in the limit of random realizations based on absorbers. 
For illustration, the posteriors for six realizations are shown in Figure~\ref{fig:realizations}. For reference we 
also plot the measurement from the full dataset as a black dot. 

\begin{figure}
\centering
\plotone{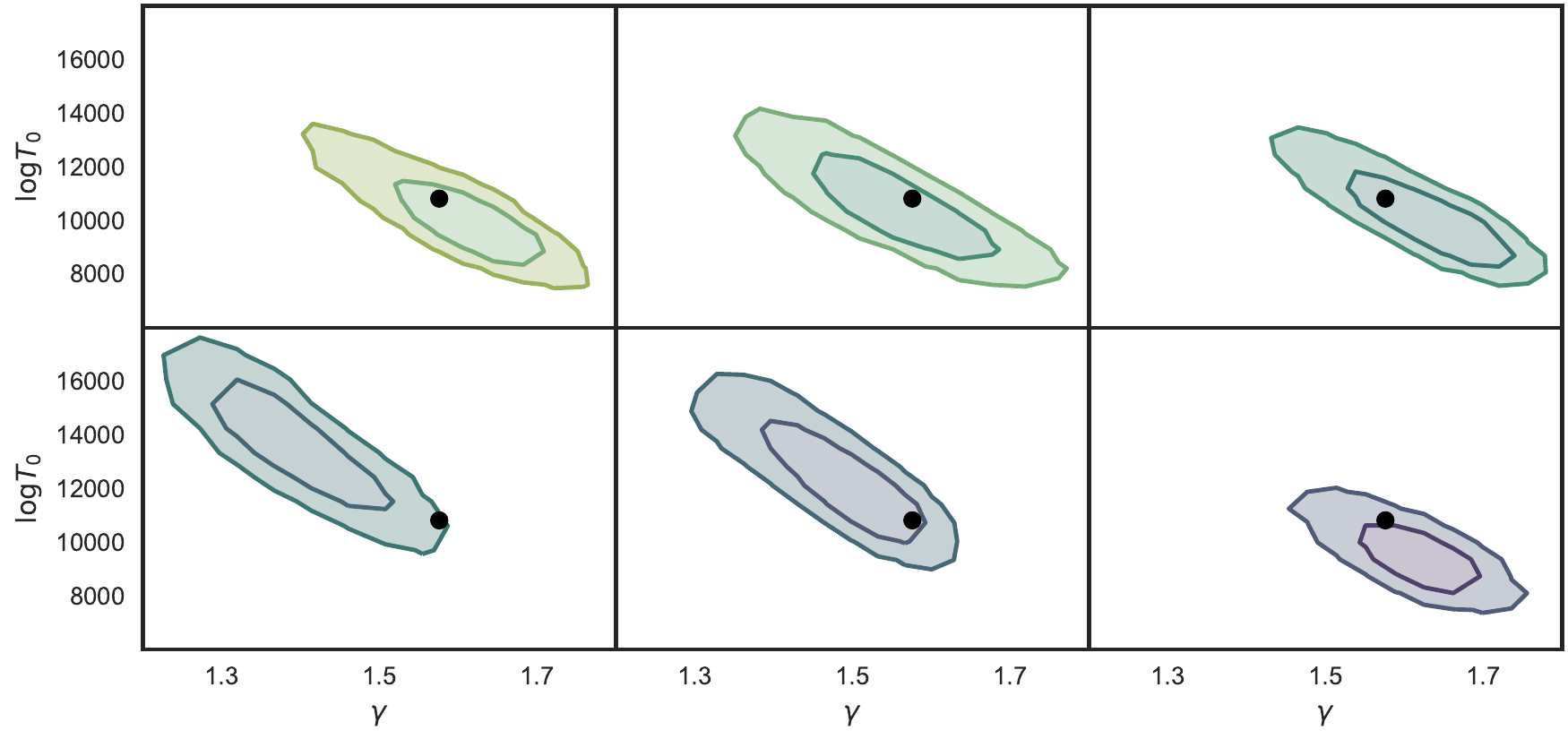}
\caption{The 1 and 2$\sigma$ contours for 6 measurement realizations (out of 200), 
each consisting of 200 unique randomly chosen absorbers from the dataset of \citet{Hiss2018} at $z=2$. The black dot illustrates the median of the measurements using the full dataset.}
\label{fig:realizations}
\end{figure}

We ran a similar test, this time choosing random QSO sightlines instead of random absorbers. 
Due to metal line masking, at $z=2$ each QSO in our sample contributes with $\sim100$ absorbers, which means that we would
carry out a measurement using around 2 sightlines each time (see discussion in \S~\ref{subsec:mock_measurement_example}). 
To test if we achieve results that are consistent with the full dataset, we carried out this experiment 
using 11 quasars that span or nearly span the pathlength within $1.9<z<2.1$, which results in 55 unique pairs of quasars 
and therefore measurement realizations. 
We observe that the reference values measured using the full dataset are within the 1$\sigma$ contours of the 2D MCMC posteriors of 
these realizations about 33\% (18/55) 
of the time. 
This implies that there is some bias associated with choosing QSO sightlines 
randomly instead of absorbers. 
Note that we do not have sufficient statistics to quantify the 
behavior of the 95\% contours with a sample size of 55 realizations. 

One possible reason for failing this inference test when choosing the pairs of QSO sightlines is the fact that we are choosing non-representative 
SNR values by picking random QSOs and comparing their \bndist s to models that were constructed to match the SNR 
distribution of the whole dataset. 
The proper approach to remove a possible SNR bias would be to generate a set of models with the matching SNR for each 
data subsample separately, i.e. generate a set of forward-models for every quasar pair in the example above. 
This approach would require applying VPFIT to our full model grid and recreating a \bndist{} emulator for every MCMC posterior we wish to generate. We have considered this approach, but concluded it implies a significant computational effort, given that the 
current calculations are already extremely resource consuming when done once. 
Additionally, real physical sightline to sightline variations in the \ac{TDR}, could also perform the poor performance on this inference
test. If present these variations would mean that subsampling by choosing random absorbers 
essentially results in a measurement of the average \ac{TDR} in that specific sub sample. 
}

\begin{acronym}
 \acro{UVB}{ultraviolet background}
 \acro{DM}{dark matter}
 \acro{UVES}{Ultraviolet and Visual Echelle Spectrograph}
 \acro{HIRES}{High Resolution Echelle Spectrometer}
 \acro{KDE}{Kernel Density Estimation}
 \acro{IGM}{intergalactic medium}
 \acro{MCMC}{Markov Chain Monte Carlo}
 \acro{PDF}{probability density function}
 \acro{PCA}{principal component analysis}
 \acro{PKP}{\ac{PCA} decomposition of \ac{KDE} estimates of a \ac{PDF}}
 \acro{UV}{ultraviolet}
 \acro{SNR}{signal-to-noise ratio}
 \acro{THERMAL}{Thermal History and Evolution in Reionization Models of Absorption Lines}
 \acro{TDR}{temperature-density relation}
 \acro{GP}{Gaussian process}
\end{acronym}

\end{document}